\documentclass[a4paper,12pt]{article}

\usepackage{psfrag}
\usepackage{mathrsfs}

\usepackage{graphicx}
\usepackage[margin=15pt,font=small,labelfont=bf,labelsep=period]{caption}
\usepackage{amsmath}
\usepackage{amssymb}
\usepackage{amsfonts}
\usepackage{mathtools}
\usepackage{bm}
\usepackage{dsfont}
\usepackage{cite}
\usepackage{setspace}
\usepackage{environ,letltxmacro}
\usepackage{tikz}
\usetikzlibrary{arrows,decorations.pathreplacing}
\usepackage{enumitem}
\usepackage{hyperref}
\usepackage{relsize}
\usepackage[a4paper,textwidth=16.04cm,textheight=22cm,footskip=15mm]{geometry}

\usepackage{empheq}
\usepackage{environ}
\NewEnviron{boxalign}{\begin{empheq}[box=\fbox]{align} \BODY \end{empheq}}
\usepackage{color}

\numberwithin{equation}{section}

\LetLtxMacro\oldequation\equation
\LetLtxMacro\endoldequation\endequation
\let\equation\relax
\let\endequation\relax
\NewEnviron{equation}[1][\empty]{%
 \ifx \empty#1
  \oldequation \BODY \endoldequation
 \else
  \ifx b#1
   \oldequation \fbox{$\displaystyle \BODY $} \endoldequation
  \else
   \oldequation \BODY \endoldequation
  \fi
 \fi}
\newcommand{\e}{\mathrm{e}}

\newcommand{\mn}{{\mu\nu}}

\newcommand{\bg}{\begin{equation}}
\newcommand{\eg}{\end{equation}}
\newcommand{\cd}{\cdots\!\,}

\def\bra{\bigl\langle}
\def\ket{\bigr\rangle}
\def\d{\mbox{d}}
\def\m{_{\mu}}
\def\n{_{\nu}}
\def\M{^{\mu}}
\def\N{^{\nu}}
\def\mn{_{\mu\nu}}
\def\MN{^{\mu\nu}}

\def\e{\mbox{e}}

\begin{document}

\begin{titlepage}

\title{
\begin{flushright}
\normalsize{MITP/18-103}
\bigskip
\vspace{1cm}
\end{flushright}
Geometric operators in the asymptotic safety scenario for quantum gravity \\[2mm]
}

\date{}

\author{Maximilian Becker and Carlo Pagani\\[3mm]
{\small Institute of Physics, 
Johannes Gutenberg University Mainz,}\\[-0.2em]
{\small Staudingerweg 7, D--55099 Mainz, Germany}
}

\maketitle
\thispagestyle{empty}

\vspace{2mm}
\begin{abstract}

We consider geometric operators, such as the geodesic length and the volume of hypersurfaces, in the context
of the Asymptotic Safety scenario for quantum gravity.
We discuss the role of these operators from the Asymptotic Safety perspective,
and compute their anomalous dimensions within the Einstein-Hilbert truncation.
We also discuss certain subtleties arising in the definition of such geometric operators.
Our results hint to an effective dimensional reduction of the considered geometric operators.

\end{abstract}

\end{titlepage}

\newpage

\begin{spacing}{1.1}

\section{Introduction}
\indent

In the Asymptotic Safety (AS) scenario, quantum gravity is described
by a path integral which is well-defined thanks to the presence of
an ultraviolet non-Gaussian fixed point having a finite
number of relevant directions \cite{W80,R98}. As such, most
of the investigations in this setting have focussed on the presence
of a suitable fixed point. The framework which is employed to test
the presence of such a fixed point is the functional renormalization
group (FRG) based on the so called effective average action (EAA)
\cite{W93}, which is a scale dependent generalization of the standard
effective action that realizes the Wilsonian renormalization program.
The extension of the EAA employed to study quantum gravity has been
proposed in \cite{R98}. The FRG provides an exact functional equation
that can be studied by implementing some approximation scheme. The
presence of a suitable fixed point has been tested under many different
approximations such as the curvature expansion and the vertex expansion.
All these different approximations confirm the presence of a suitable
UV fixed point.

The aim of this work lies on a different research line than the study
of the nature of the AS fixed point. Assuming that a suitable fixed point
exists and, for all practical purposes, employing the fixed point
stemming from the Einstein-Hilbert truncation, we study the quantum
properties of geometrical objects such as the volume of hypersurfaces in spacetime
and the geodesic length. The purpose of looking into such geometric
operators is the following. 
As discussed in \cite{Pagani:2016dof}, in
order to make contact with gravitational observables, some further
effort is needed on top of studying the renormalization of the EAA
and finding the critical exponents associated to the AS fixed point.
(This is true also for other quantities such as the entanglement entropy \cite{Pagani:2018mke}.)
Moreover, in a gravitational theory, it is rather natural to ask how
geometric quantities, like the volume of a submanifold, behave at
the quantum level. 
Indeed, even if such quantities are not true (diffeomorphism invariant) observables, 
their study may hint at the presence of certain general features
of the underlying quantum theory, such as the dynamical dimensional reduction, 
which has been already observed via different criteria and means in the literature \cite{Lauscher:2005qz,Reuter:2011ah,Carlip:2017eud}.

The paper is organized as follows. 
In section \ref{sec:AS-and-geom-oper} we first present some general considerations regarding
the geometric operators that we consider in this work. 
Then we introduce our general framework by discussing the gravitational EAA and explaining how to take
into account geometric operators.
In section \ref{sec:vol-hypersurfaces} we study the volume of arbitrary hypersurfaces and their scaling properties.
Section \ref{sec:geod-length} is devoted to the study of the geodesic length.
We first emphasise the role of different possible initial or boundary conditions defining the geodesic length 
and then we discuss the scaling properties accordingly.
We summarize our findings in section \ref{sec:summary}.

\section{Asymptotic Safety and geometric operators} \label{sec:AS-and-geom-oper}

\subsection{Geometric operators in Asymptotic Safety \label{sub:Geometric-operators-in-AS}}
\indent

The core idea of the AS program is to provide a well-defined path
integral over geometries. Ultimately, the purpose is to use such a path
integral to make predictions for the relevant quantum gravitational
observables. A particularly interesting example is that of fixed geodesic
length correlation functions of two operators, which is defined by
(see e.g.~\cite{Ambjorn:1997di,Hamber:2009zz})
\begin{eqnarray}
G\left(r\right) & = & \Bigr\langle\frac{1}{\rm{Vol}}\int\mathrm d^{d}x\sqrt{g\left(x\right)}\int\mathrm  d^{d}y\sqrt{g\left(y\right)}\,O_{1}\left(x\right)O_{2}\left(y\right)\delta\left(r-\ell_{g}\left(x,y\right)\right)\Bigr\rangle\,
\label{eq:def-correlation-length-fixed-geod-dist}
\end{eqnarray}
where $\ell_{g}\left(x,y\right)$ is the geodesic length between $x$ and $y$.

In the fixed point regime, where scale invariance is realized, one
expects $G\left(r\right)\propto r^{\alpha}$. The question is then
how to predict the scaling exponent $\alpha$. Scaling arguments
regarding correlation functions such as $G\left(r\right)$ have been
developed, and particular attention has been paid to the two dimensional
case \cite{KPZ88,DDK88}, see \cite{Reuter:1996eg,Codello:2014wfa,Pagani:2016dof} for a FRG perspective.
 
The crucial point which we would like to emphasize here is that, in order to perform
analogous scaling reasonings, it is essential not only to know the
scaling behaviour of operators such as $\sqrt{g\left(x\right)}O_{1}\left(x\right)$,
which may be characterized by a critical exponent, but also of the
geodesic length $\ell_{g}\left(x,y\right)$, which is typically not contained
in the EAA as we shall argue in section \ref{sub:Asymptotic-Safety-truncations}.

Besides correlation functions like $G\left(r\right)$, in a quantum
gravitational theory one may be interested also in further geometric
properties of spacetime. For instance, one can ask about the effective
Hausdorff dimension of the spacetime, or the spectral and the walk
dimensions. These latter quantities have been estimated in the AS
scenario in \cite{Lauscher:2005qz,Reuter:2011ah,Rechenberger:2012pm,Calcagni:2013vsa}. 
As far as the Hausdorff dimension $d_{{\rm H}}$
is concerned, it turns out that it coincides with the topological
dimension $d$ of the manifold, i.e.~$d_{{\rm H}}=d$. We shall briefly
come back to this point in section \ref{sub:vol-geod-ball}, by extrapolating the Hausdorff
dimension from the scaling behaviour of a geodesic ball of radious
$r$ in the limit for $r\rightarrow0$. Note that this definition
of the Hausdorff dimension involves geometric quantities, such as the volume of
a geodesic ball, that are again difficult to access within the standard
gravitational EAA approach.

The spectral dimension and the kinematical properties of the propagator
hint at an effective dimensional reduction near the fixed point
\cite{Reuter:2011ah}, see \cite{Carlip:2017eud} for a general discussion regarding dimensional
reduction in quantum gravity. One may thus wonder if there are other
signatures of a similar dimensional reduction when approaching the
fixed point. A possibly simple way is the following.

Let us consider a generic hypersurface in Euclidean spacetime characterized by a scale $L$.
Classically, if this hypersurface has dimension $d_{\sigma_n}$, one
expects that the volume of the hypersurface, denoted $V_{\sigma_n}$,
scales as $\langle V_{\sigma_n}\rangle\propto L^{d_{\sigma_n}}$. At the
quantum level, one expects corrections to the classical scaling though.
In section \ref{sec:vol-hypersurfaces} we will provide a first estimate of
such corrections. As we shall see, they hint again at an effective dimensional reduction
when approaching the AS fixed point.

\subsection{The gravitational EAA \label{sub:Asymptotic-Safety-truncations}}

The EAA is defined by introducing a scale $k$ below which the integration
of momentum modes is suppressed. This is achieved by adding the term
$\Delta S_{k}=\frac{1}{2}\int\chi {\cal R}_{k}\chi$ to the bare action,
with $\chi$ being the fluctuating field and ${\cal R}_{k}$ a suitable kernel.
The scale dependence of the EAA is governed by the following exact
equation, called the functional renormalization group equation \cite{W93,Ellwanger:1993mw,Morris:1993qb},
\begin{eqnarray}
\partial_{t}\Gamma_{k} & = & \frac{1}{2}\mbox{STr}\left[\left(\Gamma_{k}^{\left(2\right)}+{\cal R}_{k}\right)^{-1}\partial_{t}{\cal R}_{k}\right]\label{eq:flow_eq_EAA}
\end{eqnarray}
where $\Gamma_{k}^{\left(2\right)}$ is the Hessian of the effective
average action $\Gamma_{k}$. 

In order to actually solve equation
(\ref{eq:flow_eq_EAA}) it is necessary to introduce some approximation
scheme. A possible strategy is to expand the EAA in a finite number
of monomials:
\begin{eqnarray}
\Gamma_{k} & \equiv & \sum_{i=1}^{m}g_{i}M_{i}\,,
\end{eqnarray}
where $M_{i}\in\left\{ \int\sqrt{g},\int\sqrt{g}R,\cdots\right\} $.
Such truncations have been studied in an increasing order of complexity
by including higher curvature terms \cite{Lauscher:2002sq,Reuter:2002kd,Codello:2007bd,Codello:2008vh,Benedetti:2009rx,Benedetti:2010nr,
Ohta:2013uca,Benedetti:2013jk}, 
the Goroff-Sagnotti counterterm \cite{Gies:2016con},
and polynomials of the Ricci scalar of high order \cite{Falls:2014tra,Ohta:2015efa,Falls:2016wsa,Falls:2016msz}. 
Also functional, i.e.~infinite dimensional, truncations have been investigated in various settings
\cite{Reuter:2008qx,Benedetti:2012dx,Demmel:2012ub,Dietz:2012ic,Bridle:2013sra,Dietz:2013sba,Demmel:2014sga,
Demmel:2014hla,Demmel:2015oqa,Ohta:2015fcu,Labus:2016lkh,Dietz:2016gzg,Knorr:2017mhu,Falls:2017lst,Alkofer:2018fxj}.
Along a different direction, ans{\"a}tze taking into account the bimetric character of the flow
have been studied \cite{Manrique:2009uh,Manrique:2010mq,Manrique:2010am,Becker:2014qya}.
A further possible approximation scheme which tackles the bimetric character of the EAA is the following one. 
The EAA is expanded in a series of vertices, which are defined schematically by taking functional derivatives of the EAA,
\begin{eqnarray}
\Gamma_{k}^{\left(m,n\right)} & \equiv & \frac{\delta^{m+n}\Gamma_{k}}{\delta\bar{g}^{m}\delta h^{n}}\,,
\end{eqnarray}
where $\bar{g}$ is the background metric and $h$ the average fluctuation
field. This approach also leads to a picture which is consistent with
the AS fixed point 
\cite{Christiansen:2012rx,Codello:2013fpa,Christiansen:2014raa,Christiansen:2015rva,Knorr:2017fus,Christiansen:2017bsy,Eichhorn:2018akn}.
Moreover, the AS fixed point persists also in presence of a suitable matter content, 
see e.g.~\cite{Dou:1997fg,Percacci:2002ie,Dona:2013qba,Christiansen:2017cxa,Eichhorn:2018yfc}.

To summarize, general EAA truncations are built out of (possibly infinitely many) quasilocal operators. 
It follows that such truncations do not carry any information regarding non-local operators
such as the geodesic length.
Moreover, since one has to resort to some approximation scheme it must not be given for
granted that a given scheme is suitable for all operators. Hence it
is convenient to have a framework that allows one to study a given operator
and its renormalization over and above the standard EAA renormalization.
As we shall see in section \ref{sub:Geometric-operators-as-composite-operators},
such a framework is provided by a source dependent extension of the
standard EAA. 

As far as the EAA is concerned, in the present work we shall employ the Einstein-Hilbert truncation \cite{R98}
defined in (\ref{eq:EH_truncation}) below.
In the gravitational EAA framework for metric gravity, the metric $g_{\mu\nu}$ is
expressed via the sum of a background metric $\bar{g}_{\mu\nu}$ and
the fluctuating metric $h_{\mu\nu}$, i.e.~$g_{\mu\nu}=\bar{g}_{\mu\nu}+h_{\mu\nu}$.
(Also, different parametrizations have been investigated, see e.g.~\cite{deBrito:2018jxt}.)
However, the gauge-fixing and the cutoff action break the split symmetry \cite{Manrique:2009uh} so that a generic ansatz for $\Gamma_k$
depends not only on $g_{\mu\nu}$ but also on $\bar{g}_{\mu\nu}$ (or, equivalently, on $\bar{g}_{\mu\nu}$ and $h_{\mu\nu}$).
In the present work, we will restrict ourselves to a single metric type of truncations in which the EAA is a functional depending only
on $g_{\mu\nu}=\bar{g}_{\mu\nu}+h_{\mu\nu}$ and has the form
\begin{eqnarray}
\Gamma_{k}\left[g \right] & = & \frac{1}{16\pi G_k}\int\mathrm d^{d}x\,\sqrt{g}\left(2\Lambda_k -R\right)\,. \label{eq:EH_truncation}
\end{eqnarray}
For later convenience we introduce $G_k \equiv \bar{G} / Z_{Nk}$ and $\kappa^2 \equiv \left(32 \pi G_k \right)^{-1}$.
We equip the ansatz (\ref{eq:EH_truncation}) with the Feynman-de
Donder gauge fixing, which gives a particularly simple Hessian, see
for instance \cite{Codello:2008vh}.
Moreover, we set the cutoff kernel to be
\begin{equation}
{\cal R}_k^\mathrm{grav}[g]^{\mu\nu\rho\sigma} \equiv \kappa^2 R_k^\mathrm{grav}[g]^{\mu\nu\rho\sigma}\,,
\end{equation}
where the conformal (i.e., trace) mode is treated as proposed in \cite{R98}
\begin{equation}
R_k^\mathrm{grav}[g]^{\mu\nu\rho\sigma} \equiv 
\left((\mathrm{I}-\mathrm{P}_\phi)^{\mu\nu\rho\sigma}+\frac{2-d}{2}(\mathrm{P}_\phi)^{\mu\nu\rho\sigma}\right) R_k \left(-D^2 \right)\,.
\label{eq:def-R_k}
\end{equation}
In \eqref{eq:def-R_k} we introduced the following tensors
\begin{eqnarray}
\mathrm I_{\alpha\beta\kappa\tau} &=& \frac{1}{2}( g_{\alpha\kappa} g_{\beta\tau}+ g_{\alpha\tau} g_{\beta\kappa}) \\
(\mathrm P_\phi)_{\alpha\beta\kappa\tau} &=& \frac{1}{d} g_{\alpha\beta} g_{\kappa\tau}\, ,
\end{eqnarray}
$P_\phi$ being the projector on the trace mode.
Eventually, we will also write the function $R_k \left(-D^2 \right) $ as
$R_k \left(-D^2 \right) \equiv Z_{Nk} k^2 R^{(0)}(-D^2/k^2)$
where $R^{(0)}$ determines the cutoff shape.

 In this setting the inverse Hessian of the EAA -- which is the modified propagator for gravitons -- takes the simple form
\begin{align}
\label{eq:EHHESS}
\left(\frac{1}{\kappa^2}\Gamma^{(2)}_k[g]+R_k^\mathrm{grav}[g]\right)_{\rho\sigma\alpha\beta}^{-1}=\ \  &(\mathrm{I}-\mathrm{P}_\phi)_{\rho\sigma\alpha\beta}\,a_T^{-1}(-D^2)\notag\\
&+\frac{2}{2-d}(\mathrm{P}_\phi)_{\rho\sigma\alpha\beta}a_S^{-1}(-D^2) \, .
\end{align}
Here, with $D^2=g\MN D\m D\n$ the covariant Laplacian, $a_T(-D^2)$ and $a_S(-D^2)$ are the operators that arise in the Hessian after having split the metric $h$ into a traceless and a trace part. These operators are given by
\begin{align*}
a_T^{-1}(-D^2)&=\frac{1}{Z_{Nk}}\left[-D^2+k^2 R^{(0)}(-D^2/k^2)-2\Lambda_k+\frac{d(d-3)+4}{d(d-1)}R\right]^{-1}\; ,\\
a_S^{-1}(-D^2)&=\frac{1}{Z_{Nk}}\left[-D^2+k^2 R^{(0)}(-D^2/k^2)-2\Lambda_k+\frac{d-4}{d}R\right]^{-1} \, .
\end{align*}

\subsection{Geometric objects as composite operators \label{sub:Geometric-operators-as-composite-operators}}

In this section we argue that a possible way to tackle the description
of the geometric operators mentioned in section \ref{sub:Geometric-operators-in-AS}
is to deal with them as composite operators. Thus, let us first introduce composite operators in the FRG framework.\\

{\bf\noindent (A)} Suppose that we have a generic ansatz for the EAA,
\begin{eqnarray}
\Gamma_{k} & = & \sum_{i=1}^{m}g_{i}\int\mathrm d^{d}x\,A_{i}\left(x\right)\,,\label{eq:generic-ansatz-EAA}
\end{eqnarray}
where $g_1,\cdots,g_m$ are the coupling constants. The scaling properties
of the operators $\left\lbrace A_{i}\left(x\right) \right\rbrace$ are determined by the critical
exponents $\theta_{i}$. In particular, the scaling mass dimension
$\Delta_{i}$ is given by $\Delta_{i}=d-\theta_{i}$ where quantum
correction are taken into account. Clearly, the actual scaling
operator, having scaling dimension
$\Delta_{i}$, is given by a combination of the operators $A_1(x),\cdots,A_m(x)$,
which is determined by the linearized RG flow around the fixed point.

Given the ansatz (\ref{eq:generic-ansatz-EAA}) for $\Gamma_{k}$,
one may ask if it is possible to extract the scaling of an operator
other than $A_{i}\left(x\right)$. This is possible at the expense
of introducing further sources which are conjugate to the composite
operators of interest. To see how this comes about, let us consider
the following modified generating functional:\footnote{Whenever a dot appears, as in $J\cdot\chi$, deWitt summation and integration
convention is understood, i.e.,~$J\cdot\chi=\int\mathrm d^{d}x\,J_{a}\left(x\right)\chi^{a}\left(x\right)$.}
\begin{eqnarray}
e^{W_{k}\left[J,\varepsilon\right]} & \equiv & \int{\cal D}\chi\,e^{-S\left[\chi\right]-\Delta S_{k}\left[\chi\right]+J\cdot\chi-\varepsilon\cdot O}
\end{eqnarray}
where, on top of the standard source $J$ coupled to the elementary
field, we have introduced the sources $\left\lbrace \varepsilon_{i}\left(x\right) \right\rbrace$
which are conjugate to the set of composite operators $\left\{ O_{i}\left(x\right)\right\} $,
where $i=1,\cdots,N$. By taking functional derivatives with respect
to the source $\varepsilon_{i}\left(x\right)$ one obtains information
on the insertion of a composite operator, in particular
\begin{eqnarray}
\langle O_{i}\left(x\right)\rangle_J & = & -\frac{\delta W_{k}[J,\varepsilon]}{\delta\varepsilon_{i}\left(x\right)}\Bigg|_{\varepsilon=0}\,.
\end{eqnarray}
The functional $W_{k}\left[J,\varepsilon\right]$ is then used to
defined a generalized EAA $\Gamma_{k}\left[\varphi,\varepsilon\right]$,
which depends on the average field $\varphi$ and on the sources $\varepsilon$
(on which no Legendre transform has been performed). The insertion of a composite operator
is then defined by taking a single functional derivative with respect to the associated
source:
\begin{eqnarray}
\left[O_{k}(x)\right]_{i} & \equiv & \frac{\delta}{\delta\varepsilon_{i}(x)}\Gamma_{k}\left[\varphi,\varepsilon\right]\Bigr|_{\varepsilon=0}
\, =\, \langle O_{i}\left(x\right)\rangle_{J[\varphi]} \,,
\end{eqnarray}
where $k$ indicates the RG scale and the subscript $i$ labels the
$N$ different composite operators that we take into account.

We can write the flow equation for composite operators as \cite{Pawlowski:2005xe,Igarashi:2009tj,Pagani:2016pad,Pagani:2017tdr}
\begin{eqnarray}
\partial_{t}\left(\varepsilon\cdot\left[O_{k}\right]\right) 
& = & 
-\frac{1}{2}\mbox{Tr}\left[\left(\Gamma_{k}^{\left(2\right)}+{\cal R}_{k}\right)^{-1}\left(\varepsilon\cdot\left[O_{k}\right]^{\left(2\right)}\right)\left(\Gamma_{k}^{\left(2\right)}+{\cal R}_{k}\right)^{-1}\partial_{t}{\cal R}_{k}\right]\,.\label{eq:flow_eq_composite_operator_epsilon}
\end{eqnarray}
Again, to concretely solve equation (\ref{eq:flow_eq_composite_operator_epsilon})
some approximation scheme must be implemented. In particular, one
may expand the composite operator $\left[O_{k}\right]_{i}$ on the truncated
set of operators $\left\{ O_1,\cdots,O_N\right\} $, i.e.
\begin{eqnarray}
[O_k (x)]_{i} & = & \sum_{j=1}^{N}Z_{ij}(k)O_{j}(x) \, .
\label{eq:ansatz_composite_oper_Z_Oi}
\end{eqnarray}
Given the ansatz (\ref{eq:ansatz_composite_oper_Z_Oi}), one can show
that the scaling operators of the theory have dimension, quantum corrections
included, given by the eigenvalues of the matrix \cite{Pagani:2016pad}
\[
d_{i}\delta_{ij}+\left(Z^{-1}\partial_{t}Z\right)_{ij}
\]
where $d_{i}$ is the (classical) mass dimension of the operator $O_{i}$.

The matrix $\gamma_{Z,ij}\equiv\left(Z^{-1}\partial_{t}Z\right)_{ij}$
can be directly found manipulating equation (\ref{eq:flow_eq_composite_operator_epsilon}).
Inserting the ansatz (\ref{eq:ansatz_composite_oper_Z_Oi}) and taking
a functional derivative with respect to $\varepsilon_{i}$, we find
\begin{eqnarray}
\sum_{j}\partial_{t}\left(Z_{ij}O_{j}\left(x\right)\right) & = & -\frac{1}{2}\mbox{Tr}\left[\left(\Gamma_{k}^{\left(2\right)}+{\cal R}_{k}\right)^{-1}\left(\sum_{j}Z_{ij}O_{j}^{\left(2\right)}\left(x\right)\right)\left(\Gamma_{k}^{\left(2\right)}+{\cal R}_{k}\right)^{-1}\partial_{t}{\cal R}_{k}\right] \nonumber\\
&\,&
\end{eqnarray}
which implies
\begin{eqnarray}
\sum_{j}\gamma_{Z,ij}O_{j}\left(x\right) & = & -\frac{1}{2}\mbox{Tr}\left[\left(\Gamma_{k}^{\left(2\right)}+{\cal R}_{k}\right)^{-1}\left(O_{i}^{\left(2\right)}\left(x\right)\right)\left(\Gamma_{k}^{\left(2\right)}+{\cal R}_{k}\right)^{-1}\partial_{t}{\cal R}_{k}\right]\,.\label{eq:flow_eq_gamma_Z}
\end{eqnarray}

In this work we shall employ non-mixing ans\"{a}tze for the composite
operators. This means that we shall consider composite operators approximated
by the simple parametrization $\left[O_{k}\right]=Z_{O}\left(k\right)O$.
In this approximation, assuming that $[O_k]$ is independent of the ghost fields,
the anomalous dimension is given by $\gamma_O\equiv\gamma_{Z_O}\equiv Z_{O}^{-1}\partial_{t}Z_{O}$ and can be read off from 
\begin{align}
\label{eq:FRGE4}
{\gamma_O}(k)\, O[g](x)=&
-\frac{1}{2}\mathrm{Tr}
\Bigg[
\left(\frac{1}{\kappa^2}\Gamma^{(2)}_k[g]+R_k^\mathrm{grav}[\bar g]\right)_{\rho\sigma\alpha\beta}^{-1}
\frac{1}{\kappa^2}\, O^{(2)}[g](x)^{\alpha\beta\kappa\tau} \notag\\
&\times\left(\frac{1}{\kappa^2}\Gamma^{(2)}_k[g]+R_k^\mathrm{grav}[\bar g]\right)_{\kappa\tau\mu\nu}^{-1}
\left(\partial_t R_k^\mathrm{grav}[\bar g]^{\mu\nu\rho\sigma}\right)\Bigg]\, .
\end{align}

{\bf\noindent (B)} After this general introduction, let us turn to the use of composite
operators in the present work. As an example, let us consider the
case of the volume of an $n$-dimensional hypersurface $\sigma_n$ which is defined
by
\begin{eqnarray}
V_{\sigma_n} & = & \int_{\sigma_n}\mathrm d^{n}u\,\sqrt{x^\ast\!g}\,.
\end{eqnarray}
Here, $x^\ast\!g$ is the pullback of the spacetime metric $g$ onto the surface $\sigma_n$. The latter is characterized
by the map $u\mapsto x^{\mu}\left(u\right)$ and parametrized by $n$ coordinates
$u^{a}$ with $a=1,\cdots,n$ (more details can be found in section
\ref{sec:vol-hypersurfaces}). We then extend the action $S$ by including also the
term
\begin{eqnarray}
\int_{\sigma_n}\mathrm d^{n}u\,\varepsilon\left(u\right) \sqrt{x^\ast\!g} \,.
\end{eqnarray}
At the level of the source-dependent functional $\Gamma_{k}\left[\varphi,\varepsilon\right]$,
the above operator acquires a scale dependence which we parametrize
by employing the following source-dependent term:
\begin{eqnarray}
\int_{\sigma_n}\mathrm d^{n}u\,\varepsilon\left(u\right) Z_{\sigma_n}\left(k\right)\sqrt{x^\ast\!g}\,.
\end{eqnarray}
The anomalous dimension associated to the running of $Z_{\sigma_n}\left(k\right)$
is given by $\gamma_{\sigma_n}=Z_{\sigma_n}^{-1}\partial_{t}Z_{\sigma_n}$.
Recalling that the anomalous dimension $\gamma_{\sigma_n}$ adds to the mass
dimension, we have that a $n$-dimensional volume scales with an exponent
$-n+\gamma_{\sigma_n}$, rather than the classical value $-n$.

From the technical point of view, let us note that for our purposes it will be enough to consider spacetime-point independent sources
when solving equation (\ref{eq:flow_eq_composite_operator_epsilon}) since we neglect any mixing with total derivative operators. 
In our case the source $\varepsilon$ is constant and serves as a bookkeeping parameter. 

We also wish to emphasize that our work is of exploratory and qualitative nature. 
In order to reach quantitative predictions for the anomalous dimension a systematic study of increasingly complex truncations
for the EAA and for the composite operators must be implemented.
However, we believe that the present simplified treatment paves the way for a detailed study of the geometric operators in AS
and may already highlight certain relevant features of the AS
scenario for quantum gravity.

\section{Volume of hypersurfaces in spacetime} \label{sec:vol-hypersurfaces}

\paragraph{(A) Definition of the surface volume.} 
We consider the volume of an $n$-dimensional submanifold $\sigma_n$ that is embedded in the $d$-dimensional spacetime manifold $M$. 
To simplify our task, we assume that $\sigma_n$ can be covered by a single chart, in the following sense.

Let $x=(x^1,\cd,x^d):U\subset M \to U'\subset\mathbb{R}^d$ be a chart of $M$, and $u=(u^1,\cd,u^n):\tilde U\subset \sigma_n \to\tilde U'\subset\mathbb{R}^n$ a chart of $\sigma_n$. By pulling back the metric $g$ on $M$ with the inclusion map $\imath : \sigma_n \hookrightarrow M$, a metric $\imath^*g$ on $\sigma_n$ is induced. For $p\in U\cap\tilde U$ this induced metric is locally given by
\begin{align*}
g_{ab}(p)& \equiv
(\imath^*g)_p(\partial/\partial u^a|^p,\partial/\partial u^b|^p)=g\mn(p)\frac{\partial}{\partial u^a|^p}(x\M\circ\imath)\frac{\partial}{\partial u^b|^p}(x\N\circ\imath) \, .
\end{align*}

As $\sigma_n \subset M$ is an embedded submanifold, the inclusion map $\imath : \sigma_n \hookrightarrow M$ is an immersion, i.e. $\d\imath_p : T_p\sigma_n\to T_pM$ is injective for all $p\in M$. This means that we can identify the tangent space $T_p\sigma_n$ with the subset $\d\imath_p(T_p\sigma_n)\subset T_pM$. Consequently, we can also identify the canonical vector fields $(\partial/\partial u^1,\cd,\partial/\partial u^n):\tilde U\subset \sigma_n \to T\sigma_n$ as vector fields on $\tilde U$ mapping to $TM$ such that we can write
\begin{align*}
g_{ab}(p)&=g\mn(p)\,\d x\M|^p\!\left(\partial/\partial u^a|^p\right)\,\d x\N|^p\!\left(\partial/\partial u^b|^p\right)\\
&=g\mn(p)\frac{\partial x\M}{\partial u^a}\bigg|^p\frac{\partial x\N}{\partial u^b}\bigg|^p \, .
\end{align*}
Finally, we can trade the $p$-dependence by some dependence on $n$-dimensional coordinates and write
\begin{equation}
g_{ab}(u)=g\mn(x(u))\frac{\partial x\M(u)}{\partial u^a}\frac{\partial x\N(u)}{\partial u^b} \, .\label{eq:defgab}
\end{equation}

The volume $\mu[g]$ of the submanifold $\sigma_n$ restricted to $U\cap\tilde U$ can then be written as
\begin{align}
\label{eq:vol}
\mu[g] \equiv\int_{U\cap\tilde U}\omega_n=\int_{\tilde U''}\d^n u \sqrt{\det g (u)}
\end{align}
where $\tilde U''\equiv u(U\cap\tilde U)\subset\mathbb{R}^n$ and $\omega_n$ denotes the volume form given by the induced metric.\\

\noindent
\paragraph{(B) Some useful formulae.} 
The crucial element that enters the flow equation (\ref{eq:flow_eq_composite_operator_epsilon}) is the Hessian
of the composite operator. Thus, let us compute the Hessian of the volume of the $n$-dimensional submanifold $\sigma_n$.
In practice, we wish to evaluate
\begin{equation}
\bra x\,|\, \mu^{(2)}[g]^{\alpha\beta\kappa\tau} |\,y\ket=\frac{1}{\sqrt{\bar g(x)\bar g(y)}}\int_{\tilde U''}\d^n u\,\frac{\delta^2 \sqrt{\det g(u)}}{\delta g_{\alpha\beta}(z)\delta g_{\kappa\tau}(y)} \, . \label{eq:Hesvol}
\end{equation}
Therefore, we express $g\mn$ in terms of $g_{ab}$ and use the chain rule so that we can rewrite a functional derivative with respect to $g\mn$ as a functional derivative with respect to $g_{ab}$,
\begin{align*}
\frac{\delta}{\delta g\mn(y)}=&\int\d^n u\,\frac{\delta g_{ab}(u)}{\delta g\mn(y)}\,\frac{\delta}{\delta g_{ab}(u)}\notag\\
=&\int\d^nu\,\frac{\partial x\M}{\partial u^a}\,\frac{\partial x\N}{\partial u^b}\delta^{(d)}(x(u)-y)\,\frac{\delta}{\delta g_{ab}(u)} \, .
\end{align*}

Therewith, we can calculate the Hessian step by step, starting with the first functional derivative of the integrand $\sqrt{\det g (u)}$ of \eqref{eq:vol},
\begin{align*}
\frac{\delta\sqrt{\det g(u)}}{\delta g\mn(y)}&=\frac{1}{2}\frac{1}{\sqrt{\det g(u)}}\int\d^n u' \,\frac{\partial x\M}{\partial u'^a} \frac{\partial x\N}{\partial u'^b}\,\delta^{(d)}(x(u')-y)\,\frac{\delta \det(g(u))}{\delta g_{ab}(u')}\, .\notag\\
\end{align*}
Next, we can use Jacobi's formula for the variation of the determinant,
\begin{equation}
\frac{\delta \det(g(u))}{\delta g_{ab}(u')}=\mathrm{adj}^T(g(u))_{ab}\,\delta^{(n)}(u-u') 
\end{equation}
where $\mathrm{adj}(A)$ denotes the adjunct of the (square) matrix $A$ which is the transpose of its cofactor matrix. The adjunct matrix is related to the inverse matrix,\footnote{It should be emphasized that the relation \eqref{eq:defgab} cannot be inverted in the sense that $x(u)$ cannot be uniquely solved for $u$. However, the inverse of $g(u)$ is naturally well-defined as $\det g(u)\neq 0$.} that we will denote with upper indices as usual, by a factor of the determinant,
\begin{equation}
g^{ab}(u)\equiv(g(u)^{-1})_{ab}=\frac{1}{\det g(u)}\,\mathrm{adj}(g(u))_{ab} \, .
\end{equation}
As the adjunct of a symmeric matrix is symmetric as well, i.e. $\mathrm{adj}(g(u))_{ab}=\mathrm{adj}^T(g(u))_{ab}$, we can express the functional derivative of $\sqrt{\det g (u)}$ by means of the inverse $g^{ab}$,
\begin{equation}
\label{eq:firstvar}
\frac{\delta\sqrt{\det g(u)}}{\delta g\mn(y)}=\frac{1}{2}\sqrt{\det g(u)}\,\frac{\partial x\M}{\partial u^a} \frac{\partial x\N}{\partial u^b}\,g^{ab}(u)\,\delta^{(d)}(x(u)-y) \, .
\end{equation}

Next, it is straightforward to build the second functional derivative of $\sqrt{\det g (u)}$ using the product rule,
\begin{align}
\label{eq:2ndfuncder}
\frac{\delta^2\sqrt{\det g(u)}}{\delta g_{\alpha\beta}(z)\delta g\mn(y)}=&\frac{1}{2}\frac{\delta\sqrt{\det g(u)}}{\delta g_{\alpha\beta}(z)}\,\frac{\partial x\M}{\partial u^a} \frac{\partial x\N}{\partial u^b}\,g^{ab}(u)\,\delta^{(d)}(x(u)-y)\notag\\
+&\frac{1}{2}\sqrt{\det g(u)}\,\frac{\partial x\M}{\partial u^a} \frac{\partial x\N}{\partial u^b}\,\frac{\delta g^{ab}(u)}{\delta g_{\alpha\beta}(z)}\,\delta^{(d)}(x(u)-y) \, .
\end{align}
In the first term, we can use our result \eqref{eq:firstvar} for the first functional derivative while in the second term we can again use the chain rule to calculate the functional derivative of $g^{ab}$,
\begin{equation}
\frac{\delta g^{ab}(u)}{\delta g_{\alpha\beta}(z)}=\int\d^n u' \,\frac{\partial x^\alpha}{\partial u'^c}\frac{\partial x^\beta}{\partial u'^d}\,\delta^{(d)}(x(u')-z)\,\frac{\delta (g^{-1}(u))_{ab}}{\delta g_{cd}(u')}
\end{equation}
with 
\begin{align*}
\frac{\delta (g^{-1}(u))_{ab}}{\delta g_{cd}(u')}&=-g^{ae}(u)\,\frac{\delta g_{ef}(u)}{\delta g_{cd}(u')}\,g^{fb}(u)\\
&=-\left(g^{ac}(u)g^{db}(u)+g^{ad}(u)g^{cb}(u)\right)\delta^{(n)}(u-u') \, .
\end{align*}
Plugging these results into the RHS of \eqref{eq:2ndfuncder} we obtain the final result for the second functional derivative of $\sqrt{\det g (u)}$,
\begin{align}
\label{eq:2ndfuncder2}
\frac{\delta^2\sqrt{\det g(u)}}{\delta g_{\alpha\beta}(z)\delta g\mn(y)}=\frac{1}{4}&\sqrt{\det g(u)}\,\frac{\partial x^\alpha}{\partial u^c}\frac{\partial x^\beta}{\partial u^d}\frac{\partial x\M}{\partial u^a}\frac{\partial x\N}{\partial u^b}\,\delta^{(d)}(x(u)-y)\,\delta^{(d)}(x(u)-z)\notag\\
&\times\Bigg[ g^{ab}(u)g^{cd}(u)-g^{ac}(u)g^{db}(u)-g^{ad}(u)g^{cb}(u)\Bigg] \, .
\end{align}

Furthermore, the following traces will be needed:
\begin{align}
\mathrm I_{\alpha\beta\mu\nu}\frac{\partial x^\alpha}{\partial u^c}\frac{\partial x^\beta}{\partial u^d}\frac{\partial x\M}{\partial u^a}&\frac{\partial x\N}{\partial u^b}\left(g^{ab}g^{cd}-g^{ac}g^{db}-g^{ad}g^{cb}\right)\notag\\
&=\frac{1}{2}(g_{ac}g_{bd}+g_{bc}g_{ad})\left(g^{ab}g^{cd}-g^{ac}g^{db}-g^{ad}g^{cb}\right)=-n^2 \label{eq:traceO2-with-Id}
\end{align}
and
\begin{align}
(\mathrm P_\phi)_{\alpha\beta\mu\nu}\frac{\partial x^\alpha}{\partial u^c}\frac{\partial x^\beta}{\partial u^d}\frac{\partial x\M}{\partial u^a}&\frac{\partial x\N}{\partial u^b}\left(g^{ab}g^{cd}-g^{ac}g^{db}-g^{ad}g^{cb}\right)\notag\\
&=\frac{1}{d}g_{cd}g_{ab}\left(g^{ab}g^{cd}-g^{ac}g^{db}-g^{ad}g^{cb}\right)=\frac{1}{d}(n^2-2n)\, . \label{eq:traceO2-with-Pphi}
\end{align}
Note that in the traces \eqref{eq:traceO2-with-Id} and \eqref{eq:traceO2-with-Pphi} the partial derivatives ${\partial x}/{\partial u}$
combine with $\mathrm I$ and $\mathrm P_\phi$ in such a way to form a tensor depending solely on the metric $g_{ab}$
so that the trace gives just a simple number.

\noindent
\paragraph{(C) The anomalous dimension.}
In order to compute the anomalous dimension we insert the Hessian \eqref{eq:Hesvol}
into the flow equation (\ref{eq:FRGE4}). One finds
\begin{align}
\label{eq:EHanFRG2}
{\gamma_{\sigma_n}}(k)& \,\mu [g]=
-\frac{1}{2}\mathrm{Tr}\Bigg[\frac{1}{\kappa^2}\mu^{(2)}[g]^{\alpha\beta\kappa\tau}\partial_t R_k(-D^2)\notag\\
&\times \Bigg\{(\mathrm{I}-\mathrm{P}_\phi)_{\kappa\tau\alpha\beta}\,\left(a_T^{-1}(-D^2)\right)^2+(\mathrm{P}_\phi)_{\kappa\tau\alpha\beta}\frac{2}{2-d}\left(a_S^{-1}(-D^2)\right)^2\Bigg\}\Bigg]\, .
\end{align}
Using (\ref{eq:traceO2-with-Id}) and (\ref{eq:traceO2-with-Pphi}), it can be checked that
equation \eqref{eq:EHanFRG2} takes the form
\begin{align}
\label{eq:EHanFRG}
{\gamma_{\sigma_n}}(k)\,\mu [g]=\,&4\pi\bar G\int_{\tilde U''}\d^n u\sqrt{\det g(u)}\notag\Bigg[\frac{(d+1)n^2-2n}{d}\bra x(u)|\frac{\partial_t R_k(-D^2)}{a_T^2(-D^2)}\,|x(u)\ket\notag\\
&+\frac{4n-2n^2}{2d-d^2}\bra x(u)|\frac{\partial_t R_k(-D^2)}{a_S^2(-D^2)}\,|x(u)\ket\Bigg] \, .
\end{align}

Now we wish to expand the RHS of (\ref{eq:EHanFRG}) and retain only the term proportional to $\mu [g]$, 
from which we can read off the anomalous dimension.
Therefore, with $\mathcal{A}(-D^2)\equiv-D^2+k^2R^{(0)}(-D^2/k^2)-2\Lambda_k$, 
let us expand the operators appearing in \eqref{eq:EHanFRG} in powers of the scalar curvature $R$:
\begin{align*}
a_{T/S}^{-1}(-D^2)&=\frac{1}{Z_{Nk}}[\mathcal{A}(-D^2)+c_{T/S}R]^{-1}\\
&=\frac{1}{Z_{Nk}}\left(\mathcal{A}(-D^2)^{-1}-c_{T/S}\mathcal{A}(-D^2)^{-2}R\right)+O(R^2)\, ,
\end{align*}
respectively,
\begin{equation}
\left(a_{T/S}^{-1}(-D^2)\right)^2=\frac{1}{Z_{Nk}^2}\left(\mathcal{A}(-D^2)^{-2}-2c_{T/S}\mathcal{A}(-D^2)^{-3}R\right)+O(R^2) \, .
\end{equation}
Furthermore, we define 
$\mathcal{N}(-D^2)\equiv{\partial_t R_k(-D^2)}/(2Z_{Nk})$ such that we can expand the operators appearing in the matrix elements of  \eqref{eq:EHanFRG},
\begin{equation}
\partial_t R_k(-D^2)\left(a_{T/S}^{-1}(-D^2)\right)^2=\frac{2}{Z_{Nk}}\frac{\mathcal{N}(-D^2)}{\mathcal{A}(-D^2)^2}+O(R) \, .
\end{equation}
Note that this expression is $T$- and $S$-independent at order $R^0$ and thus the corresponding matrix elements in \eqref{eq:EHanFRG} coincide.

The matrix elements can be evaluated by employing the heat kernel technique, see e.g.~\cite{Codello:2008vh,Benedetti:2010nr}.
By considering the Laplace transform of $\mathcal{N}/\mathcal{A}^2$,
\begin{equation}
\left(\mathcal{N}/\mathcal{A}^2\right)(-D^2)=\int_0^\infty \d s\,\widetilde{\mathcal{N}/\mathcal{A}^2}(s)\,\e^{s D^2} \, ,
\end{equation}
we obtain
\begin{align*}
\bra x(u) | \,\left(\mathcal{N}/\mathcal{A}^2\right)(-D^2) \,| x(u)\ket &= \int_0^\infty \d s\,\widetilde{\mathcal{N}/\mathcal{A}^2}(s)\bra x(u) | \,\e^{s D^2} | x(u)\ket\\
&=\int_0^\infty \d s\,\widetilde{\mathcal{N}/\mathcal{A}^2}(s)\left(\frac{1}{4\pi s}\right)^{d/2}\left(1+\frac{s}{6}R\right)+O(R^2) \, .
\end{align*}
Here, we have used the asymptotic expansion of the untraced diagonal heat kernel,
see e.g.~\cite{Groh:2011dw}. 
Again dropping all curvature terms, we arrive at
\begin{align}
\bra x(u) | \,\frac{\partial_t R_k(-D^2)}{a_{T/S}^2(-D^2)}\,| x(u)\ket=\frac{2}{Z_{Nk}}\left(\frac{1}{4\pi}\right)^{d/2} Q_{d/2}[\mathcal{N}/\mathcal{A}^2]+O(R)
\label{eq:matrix-element-Rdot-prop2}
\end{align}
where we have introduced the $Q$-functionals defined by $Q_n[W]\equiv\int_0^\infty\d s \,s^{-n}\,\widetilde{W}(s)$.

Let us note that we computed the matrix element appearing in \eqref{eq:EHanFRG}
by exploiting the (untraced) heat kernel. We did so in order to make it explicit that we did not have to choose a particular background. 
It is straightforward though to re-obtain the expression (\ref{eq:matrix-element-Rdot-prop2}) choosing to project on a flat background.
In this latter case indeed one can rewrite the matrix element in terms of its Fourier transform
and arrive at the same result.

The $Q$-functionals can be expressed in terms of the threshold functions \cite{R98},
\begin{align*}
\Phi_n^p(w)&\equiv\frac{1}{\Gamma(n)}\int_0^\infty\d z z^{n-1}\frac{R^{(0)}(z)-z {R^{(0)}}'(z)}{{[z+R^{(0)}(z)+w]}^p}\ \ \text{and}\\
\tilde\Phi_n^p(w)&\equiv\frac{1}{\Gamma(n)}\int_0^\infty\d z z^{n-1}\frac{R^{(0)}(z)}{{[z+R^{(0)}(z)+w]}^p}\, .
\end{align*}
Introducing the dimensionless cosmological constant $\lambda_k\equiv\Lambda_k/k^2$ and the anomalous dimension of the Newton constant $\eta_N(k)\equiv-\partial_t\ln Z_{Nk}$, we can rewrite $Q_n[\mathcal{N}/\mathcal{A}^m]$ as
\begin{equation}
Q_n[\mathcal{N}/\mathcal{A}^m]=k^{2+2(n-m)}\left[\Phi_n^m(-2\lambda_k)-\frac{1}{2}\eta_N(k)\tilde\Phi_n^m(-2\lambda_k)\right] \, .
\end{equation}
This formula leads us to the final form of the truncated FRGE, on whose RHS we drop all curvature terms,
\begin{align}
\label{eq:anQ}
{\gamma_{\sigma_n}}(k)\,\mu [g]
=&\,4\pi\bar G\int_{U''}\d^n u\sqrt{\det g(u)}\left[\frac{(d+1)n^2-2n}{d}+\frac{4n-2n^2}{2d-d^2}\right]\notag\\
&\times\frac{2}{Z_{Nk}}\left(\frac{1}{4\pi}\right)^{d/2} Q_{d/2}[\mathcal{N}/\mathcal{A}^2]\notag\\
=&\,4\pi\bar G\,\mu [g]\left[\frac{(d+1)n^2-2n}{d}+\frac{4n-2n^2}{2d-d^2}\right]\notag\\
&\times\frac{2}{Z_{Nk}}\left(\frac{1}{4\pi}\right)^{d/2} k^{d-2}\left[\Phi_{d/2}^2(-2\lambda_k)-\frac{1}{2}\eta_N(k)\tilde\Phi_{d/2}^2(-2\lambda_k)\right] \, .
\end{align}

Let us note that dropping the curvature terms in the expansions of (\ref{eq:EHanFRG}) and (\ref{eq:matrix-element-Rdot-prop2})
amounts to an approximation. It corresponds to the non-mixing type of ans{\"a}tze
that we introduced in section \ref{sub:Geometric-operators-as-composite-operators}.
In this sense the neglected operators, such as $\sqrt{g\left(x\left(u\right)\right)} R\left(x\left(u\right)\right)$, 
are those that one would add in a more refined mixing truncation for the composite operators.

From equation \eqref{eq:anQ} it is straightforward to extract the anomalous dimension.
Defining the dimensionless Newton constant $g_k\equiv k^{d-2}\bar G/Z_{Nk}$, we have
\begin{align}
\label{eq:EHAnomDim}
{\gamma_{\sigma_n}}(\lambda_k,g_k)\equiv{\gamma_{\sigma_n}}(k)=\,&2\left(\frac{1}{4\pi}\right)^{\frac{d}{2}-1}\left[\frac{(d+1)n^2-2n}{d}+\frac{4n-2n^2}{2d-d^2}\right]\notag\\
&\ \times g_k\left[\Phi_{d/2}^2(-2\lambda_k)-\frac{1}{2}\eta_N(g_k,\lambda_k)\tilde\Phi_{d/2}^2(-2\lambda_k)\right] \, .
\end{align}
Equation \eqref{eq:EHAnomDim} constitutes one of our main results.

Few comments are in order.
First of all, the anomalous dimension in \eqref{eq:EHAnomDim} is obtained after applying a number of approximations.
In particular, the Einstein-Hilbert truncation for the EAA has been employed.
It follows that more refined EAA truncations may bring (possibly important) corrections to our result.
In a non-mixing truncation for composite operators, such as those employed in this work, 
one may project equation \eqref{eq:FRGE4} on flat spacetime. 
In this latter case, it is clear that the terms in the EAA up to the curvature squared are those which may play a non-trivial role,
since these terms appear in the regularized propagators in equation \eqref{eq:FRGE4}.
In view of this, and since the gravitational fixed point includes a relevant direction characterized by a term quadratic in curvature,
the extension of our calculation to include higher curvature terms is an important task that we leave for future work.

Let us recall that, within our approximations, equation \eqref{eq:EHAnomDim} encodes the RG flow of the composite operators
also away from the fixed point.
Employing the effective field theory (EFT) interpretation of the EAA \cite{Niedermaier:2006wt}, 
one notes that the composite operators have a (running) anomalous dimension also in the EFT framework. 
Note, however, that away from the fixed point (scaling) regime it is more difficult to interpret the anomalous dimension
since dimensionful quantities may be present at intermediate RG scales and make the scaling analysis more involved.



\paragraph{(D) Four dimensional Asymptotic Safety.}
We evaluate the numerical value of the anomalous dimension $\gamma_{\sigma_n}$ of \eqref{eq:EHAnomDim} in the case of $d=4$ at the Asymptotic Safety fixed point $(\lambda_\ast,g_\ast)$.
We consider two approximations.

 First, we compute the one-loop value of $\gamma_{\sigma_n}$, which can be retrieved from \eqref{eq:EHAnomDim}
by omitting the terms proportional to the anomalous dimension $\eta_N$ and by taking the leading order in a coupling expansion of the threshold functions.

Second, we compute the numerical value of $\gamma_{\sigma_n}$ in the fully fledged Einstein-Hilbert truncation.

In table \ref{table:fpgamma} we report the numerical values obtained for $\gamma_{\sigma_n}$. 
We denote by $\gamma_{\sigma_n}^{\mathrm{opt}}$ the anomalous dimension obtained using the optimized cutoff $R^{(0)}(z)=(1-z)\Theta(1-z)$ \cite{Litim:2001up},
and by $\gamma_{\sigma_n}^{\mathrm{exp}}$ the one obtained using the exponential cutoff $R^{(0)}(z;s)=sz/(\exp(sz)-1)$ at $s=1$. 
(Regarding the exponential cutoff, we numerically checked that our results exhibit a very mild $s$-dependence, i.e. the relative error in the region $s\in[0{.}7,1{.}5]$ was found to be $\sim3\%$
of the value of the anomalous dimension.)

\begin{table}[ht]
\caption{Fixed point values of $\gamma_{\sigma_n}$ for $d=4$. 
The first two columns show the one-loop result obtained via the optimized and the exponential cutoff.
The third and fourth columns display the results for the full fledged single metric Einstein-Hilbert truncation.} 
\centering
\renewcommand{\arraystretch}{1.5}
\begin{tabular}{c c c c c}
\hline\hline 
 & $\gamma_{\sigma_n}^{\mathrm{opt,1L}}(g_\ast^{\mathrm{opt,1L}})$ 
 & $\gamma_{\sigma_n}^{\mathrm{exp,1L}}(g_\ast^{\mathrm{exp,1L}})$ 
 & $\gamma^{\mathrm{opt}}_{\sigma_n} (\lambda_\ast^{\mathrm{opt}},g_\ast^{\mathrm{opt}})$ 
 & $\gamma^{\mathrm{exp}}_{\sigma_n} (\lambda_\ast^{\mathrm{exp}},g_\ast^{\mathrm{exp}})$ \\
\hline
$n=1$ & 0.0682 & 0.0671 & 0.0997 & 0.1006\\
$n=2$ & 0.5455 & 0.5368 & 0.7973 & 0.8044 \\
$n=3$ & 1.4318 & 1.4091 & 2.0930 & 2.1116  \\ [1ex]
\hline
\end{tabular}
\label{table:fpgamma}
\end{table}

By looking at the values in table \ref{table:fpgamma} we deduce that the anomalous dimension of a surface ${\sigma_n}$
is positive and it grows with $n$.
Remarkably, the fact that the anomalous dimension is positive implies an {\it effective dimensional reduction at the UV fixed point}.

Indeed, as anticipated in section \ref{sub:Geometric-operators-as-composite-operators},
the mass scaling dimension of $\sigma_n$ is given by $-n+\gamma_{\sigma_n}$, which corresponds to a lowered effective dimension 
$d_{\sigma_n}^{\rm{eff}}=n-\gamma_{\sigma_n}$.

As already mentioned,
in the Asymptotic Safety scenario an effective dimensional reduction has already been observed in other contexts and with a slightly different meaning.
In particular, predictions for the spectral and the walk dimensions of the whole manifold have been put forward, 
indicating an effective reduction to two dimensions in the UV limit \cite{Lauscher:2005qz,Reuter:2011ah}. 
At the same time, however, the Hausdorff dimension is still equal to the topological dimension, i.e.~$d_{\rm H}=4$ \cite{Reuter:2011ah}. 
In this section we estimated a kind of effective scaling dimension of hypersurfaces (submanifolds).
Despite our rough approximations, 
our results hint consistently that an effective dimensional reduction is indeed a general feature of the Asymptotic Safety scenario.
It should be noted also that the phenomenon of dimensional reduction is present in several quantum gravity models~\cite{Carlip:2017eud}.

Let us emphasize that either the use of the one-loop or the full-fledge Einstein-Hilbert truncation
constitute still a rather crude approximation.
In fact these truncations do not take into account the bimetric nature of the gravitational EAA \cite{RSBook}.
More precise values of the anomalous dimension $\eta_N$ can be obtained in more refined truncations, 
see \cite{Percacci:2017fkn} for an overview.
It turns out that in these schemes, at least for pure gravity, the anomalous dimension $\eta_N$ is smaller than its single metric absolute value of $2$
so that one may suspect the numerical one-loop value not to be an unreasonable approximation in this respect.
(This can be checked explicitly using some results from the literature but we leave a more complete analysis for future work.)

Furthermore, one can express the scaling of the hypersurfaces not only via the externally prescribed length $L$, 
but also via the different geometrical entities, such as the length of a given curve.
To do so, let us consider the following argument.
Since the volume $V_{\sigma_n}$ scales like $L^{n-\gamma_{\sigma_n}}$ 
and the geometric length $\ell=V_{\sigma_1}$ scales like $L^{1-\gamma_{\sigma_1}}$,
one also obtains the scaling relation
$
 V_{\sigma_n}\propto \ell^{\frac{n-\gamma_{\sigma_n}}{1-\gamma_{\sigma_1}}} 
$.


It would be interesting to generalize the above analysis to other types of theories of gravity that have shown compatibility 
with the Asymptotic Safety program, 
such as the first order formalism \cite{Daum:2010qt,Daum:2013fu},
extended theories of gravity \cite{Pagani:2013fca,Pagani:2015ema,Reuter:2015rta},
and theories on foliated spacetimes \cite{Manrique:2011jc,Rechenberger:2012dt,Biemans:2016rvp,Biemans:2017zca,Houthoff:2017oam}.

Another interesting extension of our work consists in the analysis of the flow equation for composite operators away from $d=4$.
Being a detailed analysis in this case outside the scope of the present work,
we limit ourselves to observe that in dimension $d=3$ the one-loop anomalous dimension of the parametrized hypersurfaces
takes a particular simple form, namely~$\gamma_{n}^\mathrm{1L}\stackrel{d=3}{=} \frac{n \left(n-1 \right)}{5}$,
which is independent of the cutoff profile.

\section{The geodesic length} \label{sec:geod-length}

As explained in section \ref{sub:Geometric-operators-in-AS} the geodesic length is extremely useful in defining certain
diffeomorphism invariant observables. 
In this section we study the geodesic length from the composite operator point of view and explore in particular its scaling behavior in Asymptotic Safety.

\subsection{Selecting a geodesic} \label{sec:def-geod-length}

For a given Euclidean spacetime metric $g$, the geodesic length $\ell_g$ is defined by
\begin{equation}
\ell_g \equiv \int_{0}^{1} \d\tau\,
\sqrt{g_{\mu\nu}\left(x_g\left(\tau\right)\right)\dot{x}_g^{\mu}\left(\tau\right)\dot{x}_g^{\nu}\left(\tau\right)}\,,
\label{eq:def-geod-distance}
\end{equation}
where $x_g\left(\tau\right)$, $\tau\in [0,1]$, is a solution of the geodesic equation:
\begin{equation}
\ddot{x}_g^{\mu}\left(\tau\right)+\Gamma_{\alpha\,\,\,\,\,\beta}^{\,\,\,\,\,\mu}\left(x_g(\tau)\right)\,\dot{x}_g^{\alpha}\left(\tau\right)\dot{x}_g^{\beta}\left(\tau\right)
=	0 \,. \label{eq:geodesic-equation}
\end{equation}
Clearly, the geodesic $x_g\left(\tau\right)$ is fully defined only when equation (\ref{eq:geodesic-equation}) 
is equipped with suitable initial or boundary conditions.
As we shall see, such a choice plays a major role at the quantum level and one must carefully address the different possibilities separately.
Let us list three options for the supplementary conditions with which to equip equation (\ref{eq:geodesic-equation}).
\begin{description}
\item[Boundary value problem.] The solution of equation (\ref{eq:geodesic-equation}) is fixed by requiring that the geodesic passes
through the point $x_{0}^{\mu}$ at the initial ``time'' $\tau =0$ and through the point $x_{1}^{\mu}$ at the final time $\tau =1$; thereby the points $x_{0}^{\mu}$ and $x_{1}^{\mu}$ are externally prescribed:
\begin{equation}
\begin{cases}
 x^{\mu}\left(0\right)=x_{0}^{\mu}\\
 x^{\mu}\left(1\right)=x_{1}^{\mu} \, .
\end{cases}	 \label{eq:def-boundary-value-prob}	
\end{equation}
In this case $\ell_g\equiv\ell_g(x_1,x_2)$ is referred to as the geodesic distance of the two points.\footnote{Here and in the following we assume that $x_1$ and $x_2$ are sufficiently close so that no caustics occur and the solution to \eqref{eq:geodesic-equation} with \eqref{eq:def-boundary-value-prob} is unique.}
\item[Initial value problem.] 
The solution of equation (\ref{eq:geodesic-equation}) is fixed by requiring that the geodesic passes
through the point $x_{0}^{\mu}$ at the initial time $\tau =0$ with a certain ``velocity'' $v_{0}^{\mu}$, with $x_{0}^{\mu}$ and $v_{0}^{\mu}$ externally prescribed:
\begin{equation}
\begin{cases}
 x^{\mu}\left(0\right)=x_{0}^{\mu}\\
 \dot{x}^{\mu}\left(0\right)=v_{0}^{\mu}\, .
\end{cases}   \label{eq:def-initial-value-prob}
\end{equation}
Now the final point, $x_{g}^{\mu}(1)$ arises as the result of actually solving \eqref{eq:geodesic-equation} with initial data $(x_0,v_0)$.

\item[Normalized initial value problem at fixed geodesic length.] 
The solution of equation (\ref{eq:geodesic-equation}) is fixed by requiring that the geodesic passes
through a prescribed point $x_{0}^{\mu}$, at the initial time $\tau =0$, 
and that its initial direction is parallel to an externally given normalized ``velocity'' vector $\xi_{0}^{\mu}$. 
Being the velocity vector normalized, we still need to impose one further condition.
In particular one can require that the geodesic length itself equals a prescribed value $r$:
\begin{equation}
\begin{cases}
 x^{\mu}\left(0\right)=x_{0}^{\mu}\\
 \frac{\dot{x}^{\mu}\left(0\right)}{\left\Vert \dot{x}_0 \right\Vert }=\xi_{0}^{\mu} \\
 \ell_g(x_0,x_g(1)) =r \, .
\end{cases}  \label{eq:def-initial-normalized-value-prob}
\end{equation}
\end{description}

Each of these choices provides a set of $2d$ conditions which select a unique solution of the geodesic equation (\ref{eq:geodesic-equation}).

We shall focus our attention to the condition (\ref{eq:def-boundary-value-prob}) mostly
since the geodesic distance of points appears in many correlation functions of considerable physical interest, such as (\ref{eq:def-correlation-length-fixed-geod-dist})
for example.\footnote{
In fact, the initial point $x$ and the end point $y$ appearing in $\ell_g \left(x,y \right)$ in \eqref{eq:def-correlation-length-fixed-geod-dist}
are taken as given numbers, 
i.e.~they are {\it independent} of the metric and give rise to no graviton vertex.}
Interestingly, boundary conditions involving a fixed geodesic length instead, 
as in the case of (\ref{eq:def-initial-normalized-value-prob}),
have been used in the literature to define correlators $\langle \phi \left(x\right) \phi \left(y\right)\rangle $
of a kind different from \eqref{eq:def-correlation-length-fixed-geod-dist},
see e.g.~\cite{Tsamis:1989yu,Frob:2017apy}.

\subsection{Anomalous dimension of the geodesic length}

In this section we discuss the anomalous dimension associated to \eqref{eq:def-geod-distance}.
As a preliminary observation, we note that the length of an arbitrary one dimensional curve, corresponding to (\ref{eq:vol}) for $n=1$,
is given by an integral fully analogous to \eqref{eq:def-geod-distance}.
There is, however, a crucial difference between the case of an arbitrary curve and that of the geodesic distance. 
In the former case the curve, say $x^\mu \left(u \right)$, is arbitrary and taken to be independent of the underlying metric.
On the contrary, in the latter case, the geodesic $x^\mu_g \left(\tau \right)$ depends functionally on the metric, 
as is obvious from the geodesic equation. 
Therefore, in the case of the geodesic length, there are further gravitational vertices to be taken into account. They are caused by the fact that the curve under consideration depends non-trivially
on the metric.

We now proceed to compute the novel contributions to the Hessian of $\ell_g$. 
We choose to work on a flat background, $\bar{g}_{\mu\nu}=\delta_{\mu\nu}$,
and
express the geodesic trajectory as a functional series expansion in the fluctuating metric $h_{\mu\nu}$,
\begin{equation}
x_g^{\mu}\left(\tau\right) =
x_{0}^{\mu}\left[h^0\right] \left(\tau\right)+x_{1}^{\mu} \left[h \right] \left(\tau\right)
+\frac{1}{2}x_{2}^{\mu}\left[h^2 \right]\left(\tau\right) +\cdots \,. \label{eq:formal-h-expansion-of-geodesic-sol}
\end{equation}

For our purposes it is enough to compute this expansion up to the second order in $h\mn$, since we will eventually set $h_{\mu\nu}=0$
in the flow equation. 
Such choice of background and fluctuation fields implies that the LHS of the flow equation (\ref{eq:FRGE4}) 
is proportional to 
\begin{equation}
\ell_g \Bigr|_{\bar{g}=\delta, h=0} = \int_{0}^{1} \d\tau\,
\sqrt{\delta_{\mu\nu}\dot{x}_0^{\mu}\left(\tau\right)\dot{x}_0^{\nu}\left(\tau\right)}
= \sqrt{\delta_{\mu\nu}\dot{x}_0^{\mu} \dot{x}_0^{\nu} } 
\end{equation}
where we exploit the fact that $\dot{x}_0^{\mu}\left(\tau\right)$ does not actually depend on $\tau$, 
as we shall see in a moment.
We also note that $\ell_g \propto \left\Vert \dot{x}_0 \right\Vert$ in this case.
The RHS of the flow equation (\ref{eq:FRGE4}) will generate also terms different from $\ell_g$.
We will neglect such terms and select only those that can be matched with our ansatz on the LHS.
These latter terms can then be identified
by looking at which terms are proportional to $\left\Vert \dot{x}_0 \right\Vert$ on the RHS.

To calculate the Hessian, we wish to keep orders up to $O\left(h^2\right)$, which is why we expand the connection in the geodesic equation as
\begin{equation}
\ddot{x}_g^{\mu}\left(\tau\right)+\left(\Gamma_{\alpha\,\,\,\,\,\beta}^{\,\,\,\,\,\mu}\Bigr|_{\bar{g}=\delta}+\delta\Gamma_{\alpha\,\,\,\,\,\beta}^{\,\,\,\,\,\mu}\Bigr|_{\bar{g}=\delta}+\frac{1}{2}\delta^{2}\Gamma_{\alpha\,\,\,\,\,\beta}^{\,\,\,\,\,\mu}\Bigr|_{\bar{g}=\delta}\right)\dot{x}_g^{\alpha}\left(\tau\right)\dot{x}_g^{\beta}\left(\tau\right)	=	0\, .
\label{eq:geod-eq-expanded-connection}
\end{equation}

We now proceed to solve the geodesic equation \eqref{eq:geod-eq-expanded-connection} order by order in $h$,
taking the boundary conditions \eqref{eq:def-boundary-value-prob} into account.
At zeroth order the equation simply reads 
\begin{equation}
\ddot{x}_{0}^{\mu}\left(\tau\right) = 0
\end{equation}
whose solution is
\begin{equation}
x_{0}^{\mu}\left(\tau\right) = x_{0}^{\mu}+\xi_{0}^{\mu}\tau\,. \label{eq:zero-order-geodesic-sol}
\end{equation}
Here, one has $\xi_{0}^{\mu} \equiv x_{1}^{\mu}-x_{0}^{\mu}$ so that the boundary conditions \eqref{eq:def-boundary-value-prob}
are already taken into account at order $O(h^0)$.
In turn, this implies that the higher order corrections to \eqref{eq:zero-order-geodesic-sol}, 
which appear in \eqref{eq:formal-h-expansion-of-geodesic-sol},
satisfy $x_i\left(0\right) = x_i\left(1\right) =0$.

Next, the equation for $x_1\left( \tau \right)$ takes the form
\begin{equation}
\ddot{x}_{1}^{\mu}\left(\tau\right)+\delta^{\mu\nu}
\left[\partial_{\alpha}h_{\beta\nu}\left(x_{0}\left(\tau\right)\right)+\partial_{\beta}h_{\alpha\nu}\left(x_{0}\left(\tau\right)\right)
-\partial_{\nu}h_{\alpha\beta}\left(x_{0}\left(\tau\right)\right)\right]
\xi_{0}^{\alpha}\xi_{0}^{\beta} =0 \, .
\label{eq:first-order-only-geod-eq}
\end{equation}
Defining
\begin{equation}
f_{1}^{\mu}\left(\tau\right) \equiv
\delta^{\mu\nu}
\left[\partial_{\alpha}h_{\beta\nu}\left(x_{0}\left(\tau\right)\right)+\partial_{\beta}h_{\alpha\nu}\left(x_{0}\left(\tau\right)\right)
-\partial_{\nu}h_{\alpha\beta}\left(x_{0}\left(\tau\right)\right)\right]
\xi_{0}^{\alpha}\xi_{0}^{\beta}\,,
\end{equation}
one can check that the general solution of \eqref{eq:first-order-only-geod-eq} reads
\begin{equation}
x_{1}^{\mu}\left(\tau\right) = a+b\tau-\int_{0}^{\tau}\d\eta_{2}\int_{0}^{\eta_{2}}\d\eta_{1}\,f_{1}^{\mu}\left(\eta_{1}\right)\,.
\end{equation}
Implementing the boundary conditions \eqref{eq:def-boundary-value-prob} one obtains
\begin{equation}
x_{1}^{\mu}\left(\tau\right)	=	\left(\int_{0}^{1}\d\eta_{2}\int_{0}^{\eta_{2}}\d\eta_{1}\,f_{1}^{\mu}\left(\eta_{1}\right)\right)\tau
-\int_{0}^{\tau}\d\eta_{2}\int_{0}^{\eta_{2}}\d\eta_{1}\,f_{1}^{\mu}\left(\eta_{1}\right)\,.
\end{equation}

Let us note that $x_1\left(\tau\right)$, being linear in $f_1$, is proportional to $\left(\xi_0\right)^2$.
As we mentioned before, we can drop from the RHS of the flow equation (\ref{eq:FRGE4}) all terms that are not proportional to
$\left\Vert \dot{x}_0 \right\Vert^{1/2}=\left\Vert \xi_0 \right\Vert^{1/2}$.
Since the RHS of (\ref{eq:FRGE4}) is proportional to the Hessian of $\ell_g$, 
and since $\xi_0$ appears only in this Hessian, 
it is enough to identify the terms that are proportional to $\left\Vert \xi_0 \right\Vert^{1/2}$
in the Hessian of $\ell_g$.
It turns out that, since $x_1\left(\tau\right) \propto \left(\xi_0\right)^2$, the contribution to the Hessian of $\ell_g$
coming from  $x_1\left(\tau\right)$ gives rise solely to terms which we neglect within our approximation.

A very similar analysis can be performed for the contribution coming from $x_2\left(\tau\right)$.
Also, in this case the Hessian generates terms which are of higher order in $\xi_0$
and can be neglected in our approximation.

Thus the important result is that for the case of a geodesic the only relevant contributions to the Hessian of $\ell_g$ are precisely those which appeared already when we considered the length of arbitrary, prescribed curves.
This implies that, for a non-mixing ansatz for $\ell_g$, we obtain that $\gamma_{\ell_g}=\gamma_{\sigma_{n=1}}$,
whose numerical values have already been reported in table \ref{table:fpgamma}.

Nevertheless, let us emphasise that we still generally expect that $\gamma_{\ell_g}\neq \gamma_{\sigma_{n=1}}$ for more refined mixing ans{\"a}tze.
Indeed, as soon as mixing ans{\"a}tze are considered, the neglected operators on the RHS must be taken into account:
the graviton vertices due to $x_1$ and $x_2$ will play a role leading to a different anomalous dimension.
Tentatively, one may interpret the fact of having $\gamma_{\ell_g}=\gamma_{\sigma_{n=1}}$ at the level on non-mixing ans{\"a}tze
as hinting that the two anomalous dimensions are not very much different.
This, however, should be supported by a systematic enlargement of the truncation for the composite operators at hand.
We leave this task for the future.

\subsection{On different choices of initial conditions}

In this section we consider both types of \textit{initial} conditions,
i.e.~those given in \eqref{eq:def-initial-value-prob} and \eqref{eq:def-initial-normalized-value-prob}.

\paragraph{(A) Initial value problem \eqref{eq:def-initial-value-prob}.}
We rewrite the geodesic length \eqref{eq:def-geod-distance} in terms of the initial condition \eqref{eq:def-initial-value-prob}.
Recall that the integrand in \eqref{eq:def-geod-distance}, 
i.e.~$\sqrt{g_{\mu\nu}\left(x_g\left(\tau\right)\right)\dot{x}_g^{\mu}\left(\tau\right)\dot{x}_g^{\nu}\left(\tau\right)}$,
is actually a constant of motion.
Therefore we can write
\begin{eqnarray}
\ell_g &=& \int_{0}^{1}\d\tau\,\sqrt{g_{\mu\nu}\left(x\left(\tau\right)\right)\dot{x}^{\mu}\left(\tau\right)\dot{x}^{\nu}\left(\tau\right)}
\,=\, \int_{0}^{1}\d\tau\,\sqrt{g_{\mu\nu}\left(x\left(0\right)\right)\dot{x}^{\mu}\left(0\right)\dot{x}^{\nu}\left(0\right)}
\nonumber \\
&=& \sqrt{g_{\mu\nu}\left(x_0\right) v_0^{\mu} v_0^{\nu}\left(0\right)} \,,
\label{eq:geod-length-via-initial-condition}
\end{eqnarray}
where in the second line we exploited the fact that the $\tau$-integration is trivial.

We observe that, being fixed initial conditions, both $x_0^\mu$ and $v_0^\mu$ do not have any metric dependence.
Thus, the only metric dependence in $\ell_g$ is the explicit one in \eqref{eq:geod-length-via-initial-condition}.

In the present situation, the evaluation of the Hessian of $\ell_g$ and the subsequent trace calculation proceeds very much like 
in the case of an arbitrary curve.
It follows that, at the current level of accuracy, $\gamma_{\ell_g}=\gamma_{\sigma_{n=1}}$.

It must be noted that, once again, the equality $\gamma_{\ell_g}=\gamma_{\sigma_{n=1}}$ is just approximate and holds only at the level
of non-mixing ans{\"a}tze.
Actually, given that the Hessian of $\ell_g$ differs with regard to the choice of boundary conditions \eqref{eq:def-boundary-value-prob}
or initial conditions \eqref{eq:def-initial-value-prob}, also the anomalous dimension $\gamma_{\ell_g}$
depends on the type of problem one wishes to consider.

\paragraph{(B) Normalized initial value problem \eqref{eq:def-initial-normalized-value-prob}.}
In this case
we rewrite the geodesic length \eqref{eq:def-geod-distance} in terms of the initial condition \eqref{eq:def-initial-normalized-value-prob}.

This time the task is trivial because the third equation among the conditions \eqref{eq:def-initial-normalized-value-prob} tells us that 
$\ell_g=r$.
It follows that, by definition, $\ell_g$ is nothing but a given fixed number $r$, independent of the metric.
Thus, there cannot be any quantum corrections to the dimension of $\ell_g$ since it is not influenced by gravitational fluctuation at all
(equivalently, in line with our previous treatment, the Hessian of $\ell_g$ is trivial). Hence, $\gamma_{l_g}=0$ in this case.

The choice of initial conditions \eqref{eq:def-initial-normalized-value-prob} is not particularly interesting for our present purposes.
However, such initial conditions have been used in the literature to define a different type of correlation function at fixed geodesic length.
For instance, one may consider applying this choice to correlation functions of scalar fields $\langle \phi \left(x\right) \phi \left(y\right) \rangle$.
In this case, the endpoint $y$ depends on the metric whereas the geodesic distance between $x$ and $y$ does not \cite{Tsamis:1989yu,Frob:2017apy}.
This is of course very different from the boundary conditions \eqref{eq:def-boundary-value-prob}, where both 
$x$ and $y$ are fixed.

\subsection{Comment on the volume of a geodesic ball} \label{sub:vol-geod-ball}

In this section we consider the volume of a geodesic ball from the point of view of Asymptotic Safety.
Given the center of the ball, say $x_0$, the volume of a geodesic ball with radius $r$ is defined by
\begin{equation}
V_{\rm{ball}} \left(r \right) \equiv \int_B \d^dx\, \sqrt{g}\,,
\end{equation}
where the integration domain is
\begin{equation}
B \equiv \left\lbrace x: \ell_g \left(x,x_0 \right) \leq r \right\rbrace \,, \label{eq:def-domain-integration-B}
\end{equation}
with $\ell_g \left(x,x_0 \right)$ denoting the geodesic distance between $x$ and $x_0$.

In the limit of vanishing radius, the scaling of $V\left( r\right)$ can be used to extract the Hausdorff dimension 
$d_{\rm{H}}$ via \cite{Reuter:2011ah}:
\begin{eqnarray}
\lim_{r\rightarrow0}\, \Bigr\langle V_{\rm{ball}}\left(r\right) \Bigr\rangle	
& \propto &	r^{d_{\rm{H}}}\,. \label{eq:def-Hausdorff-dim-via-ball-volume}
\end{eqnarray}

As we have seen in section \ref{sec:def-geod-length}, it is crucial to state the boundary conditions that define our problem precisely.
In particular, we shall be interested in the case where the ball radius $r$ is a given {\it fixed} quantity
independent of the metric, 
i.e.~the domain of integration \eqref{eq:def-domain-integration-B} is specified by the fixed center of the ball $x_0$ and the fixed radius $r$.

To make progress, let us consider the Riemann normal coordinate expansion for the geodesic length.
Let us denote by $\xi^\mu$ the Riemann normal coordinates based at $x_0$. By definition, they correspond to the initial ``velocity'' vector of the geodesic at $x_0$.
The integration domain $B$ in \eqref{eq:def-domain-integration-B} is then rewritten as
\begin{equation}
B = \left\lbrace \xi : \sqrt{g_{\mu\nu} \left(x_0 \right)  \xi^\mu \xi^\nu}  \leq r  \right\rbrace     \,,
\end{equation}
where we used the fact that the integrand in the definition of $\ell_g$ is actually a constant of motion.
The volume element can be expanded as follows
\begin{equation}
\sqrt{g\left(\xi\right)} \approx
\sqrt{g\left(x_{0}\right)}\left\{ 1-\frac{1}{6}R\mn\left(x_{0}\right)\xi\M\xi\N\right\} \,. 
\label{eq:RNC-expansion-for-sqrtG}
\end{equation}
Thus, the volume of the geodesic ball can be rewritten as
\begin{equation}
V_{\rm{ball}} \left(r \right) = \int_B \d^d \xi \, \sqrt{g\left(\xi \right)} = 
\int_B \d^d \xi \left\lbrace \sqrt{g\left(x_0 \right)} + O\left(R, \xi \right)  \right\rbrace \,.
\label{eq:RNC-expansion-for-Vol-ball}
\end{equation}

For the time being, let us consider only the first term on the RHS of \eqref{eq:RNC-expansion-for-Vol-ball} and neglect higher order
curvature terms.
The geodesic length can be further rewritten using the vielbein $e_{\mu}^a$ given by $g\mn=\delta_{ab}\,e_{\mu}^a e_{\nu}^b$,
\begin{eqnarray}
\sqrt{g_{\mu\nu}\left(x_{0}\right)\xi^{\mu}\xi^{\nu}}	=
\sqrt{\delta_{ab}e_{\mu}^{a}\left(x_{0}\right)e_{\nu}^{b}\left(x_{0}\right)\xi^{\mu}\xi^{\nu}} 
=	\sqrt{\delta_{ab}y^{a}y^{b}}\,,
\end{eqnarray}
with $y^{a}\equiv e_{\mu}^{a}\left(x_{0}\right)\xi^{\mu}$. 
Using the $y$-coordinates it is straightforward to check that
\begin{eqnarray}
\int_{B}\d^{d}\xi\,\sqrt{g\left(x_{0}\right)}	= \frac{\pi^{d/2}}{\Gamma\left(\frac{d}{2}+1\right)}r^{d} \, ,
\label{eq:leading-ball-volume-via-RNC}
\end{eqnarray}
which is the standard result for flat space.

Since, by definition, the radius $r$ is a fixed number independent of the metric, we obtain for the expectation value of \eqref{eq:leading-ball-volume-via-RNC}:
\begin{eqnarray}
\Bigr\langle \int_{B}\d^{d}\xi\,\sqrt{g\left(x_{0}\right)} \Bigr\rangle	
= \frac{\pi^{d/2}}{\Gamma\left(\frac{d}{2}+1\right)}r^{d} \,. 
\label{eq:average-leading-ball-volume-via-RNC}
\end{eqnarray}

Higher order terms can be found in an analogous way, see \cite{Gray74} for a detailed presentation.
For instance, the second term in the brackets on the RHS of \eqref{eq:RNC-expansion-for-sqrtG} gives a contribution 
of the following kind:
\begin{equation}
\Bigr\langle \int_{B}\d^{d}\xi\,\sqrt{g\left(x_{0}\right)} R\mn\left(x_{0}\right)\xi\M\xi\N \Bigr\rangle	
\, \propto \, \Bigr\langle R\left(x_0 \right) \Bigr\rangle \, r^{d+2} \,.
\label{eq:next-leading-ball-volume-via-RNC}
\end{equation}
Contrary to the leading term \eqref{eq:leading-ball-volume-via-RNC}, we have some metric dependence in \eqref{eq:next-leading-ball-volume-via-RNC}.
However, there is no reason for $ \langle R\left(x_0 \right) \rangle $ to be a function of $r$
($ \langle R\left(x_0 \right) \rangle $ may be expected to be a function of $x_0$).
Assuming its $r$-dependence, and that similar properties are shared by the other terms in the expansion \eqref{eq:RNC-expansion-for-Vol-ball},
we limit ourselves to consider just the leading term \eqref{eq:leading-ball-volume-via-RNC}
and neglect the higher terms which are suppressed by further powers of $r$ in the limit $r\rightarrow 0$.
At the end of the section we will state the condition under which this approximation is possible and provide an argument for it.

Let us now come back to the Hausdorff dimension defined by equation \eqref{eq:def-Hausdorff-dim-via-ball-volume}.
In the limit $r\rightarrow 0$, the relevant leading term 
to deduce the scaling power in \eqref{eq:def-Hausdorff-dim-via-ball-volume}
is given by equation \eqref{eq:average-leading-ball-volume-via-RNC},
which is not affected by any anomalous scaling due to the fact that $r$ is independent of the metric.
This implies that in the EAA approach to Asymptotic Safety {\it the Hausdorff dimension of spacetime is equal to the topological dimension: $d_{\rm{H}}=d$}.
This is one of our main results. It confirms the result already obtained in \cite{Reuter:2011ah} via a different argument.

Several comments are in order here. Let us consider under which condition one is allowed to neglect the higher order curvature terms.
In order to have an idea of the behaviour of these terms approaching the UV fixed point,
let us employ a ``mean field'' kind of estimate:
\begin{equation}
\Bigr\langle R\left( x_0 \right) \Bigr\rangle \approx R\left( x_0 \right)\Bigr|_{g_{\mu\nu} =\bar{g}^{\rm{sc}}_{\mu\nu}}\, .
\end{equation}
Here $\bar{g}^{\rm{sc}}_{\mu\nu}$ is the so called ``self-consistent background metric''~\cite{Becker:2014pea}, which is a special background field configuration 
such that $\langle h\mn\rangle=0$ which implies $\langle g_{\mu\nu}\rangle = \bar{g}^{\rm{sc}}_{\mu\nu}$.
The metric $\bar{g}^{\rm{sc}}_{\mu\nu}$ is $k$-dependent, and in the UV fixed point limit it behaves as
$\bar{g}^{\rm{sc}}_{\mu\nu} \propto k^{-2}$.
It follows that in the UV fixed point regime we expect
\begin{equation}
\Bigr\langle R\left( x_0 \right) \Bigr\rangle \propto k^2 \,. \label{eq:average-R-UV-limit}
\end{equation}
Thus, in the UV limit $k\rightarrow \infty$ the average curvature blows up.

The estimate \eqref{eq:average-R-UV-limit} seems to prevent us from truncating the expansion \eqref{eq:RNC-expansion-for-Vol-ball}
to its leading term \eqref{eq:average-leading-ball-volume-via-RNC}.
However, one must note that the higher order terms are suppressed by higher powers of the radius $r$.
Let us consider the example of the next-to-leading term \eqref{eq:next-leading-ball-volume-via-RNC}, we expect
\begin{equation}
\Bigr\langle \int_{B}\d^{4}\xi\,\sqrt{g\left(x_{0}\right)} R\mn\left(x_{0}\right)\xi\M\xi\N \Bigr\rangle	
\, \propto \, \Bigr\langle R\left(x_0 \right) \Bigr\rangle \, r^{d+2} 
\, \propto \, r^d \left(kr \right)^2 \,.
\label{eq:next-leading-ball-volume-estimate}
\end{equation}
From \eqref{eq:next-leading-ball-volume-estimate} we see that our procedure is justified as long as $kr \ll 1$.
The scale $k^{-1}$ corresponds to the largest length that we have integrated out in the flow. 
Therefore the condition $kr \ll 1$ corresponds to requiring that the radius $r$ is inside the range of lengths that have
already been integrated by the flow, i.e.~$0 < r \ll k^{-1}$.
This requirement is a physical one since only in this way the ball is affected by all the relevant modes.

Our argument can not be applied straightforwardly if the estimate \eqref{eq:average-R-UV-limit} is not a good approximation.
Thus, let us provide a further argument that does not rely on \eqref{eq:average-R-UV-limit}, but which is
less physically intuitive.
In our non-mixing ansatz the operator $V_{\rm{ball}}\left(r\right)$ renormalizes multiplicatively.
The anomalous dimension is in general scale dependent: $\gamma_{V_{\rm{ball}}}= \gamma_{V_{\rm{ball}}}\left( k \right)$.
At the UV fixed point one expects $\gamma_{V_{\rm{ball}}}$ to tend to a constant in a smooth way.
Furthermore, from the numerical viewpoint, one expects that close enough to the fixed point, say at a scale $k_{\rm{big}}$,
we have $\gamma_{V_{\rm{ball}}}\left( k_{\rm{big}} \right) \approx \gamma_{V_{\rm{ball}}}\left( \infty \right)$.
For all practical purposes then, we could limit ourselves to compute $\gamma_{V_{\rm{ball}}}\left( k_{\rm{big}} \right)$
via the usual flow equation.
Now, at a finite (but large) scale $k_{\rm{big}}$, even the average curvature $\langle R \rangle$ is expected to be finite
and the higher order terms in \eqref{eq:RNC-expansion-for-Vol-ball} are expected to be negligible for $r\rightarrow 0$,
so that we could limit ourselves to consider only the leading order term \eqref{eq:leading-ball-volume-via-RNC}.

\section{Summary} \label{sec:summary}

In this work we have made the first step in addressing the study of geometric operators,
such as the volume of hypersurfaces and the geodesic length, in the context of the Asymptotic Safety scenario for quantum gravity.

In section \ref{sec:AS-and-geom-oper} we have argued that geometric operators are important quantities
which reflect characteristic features of the Asymptotic Safety scenario, and they may be used to make contact
with full fledged observables. We have also detailed our main tools and approximation schemes.

In section \ref{sec:vol-hypersurfaces} we studied the anomalous scaling of hypersurfaces.
The calculation of the anomalous dimension associated to such hypersurfaces shows an effective dimensional reduction
in the fixed point regime. 
The anomalous dimension grows with the topological dimension of the hypersurface, but the sign of the correction always implies
an effective dimensional reduction.

In section \ref{sec:geod-length} we have studied the geodesic distance and its scaling properties.
Such properties are important in making contact with the scaling of observable correlation functions,
such as that in \eqref{eq:def-correlation-length-fixed-geod-dist}.
We first noted that a crucial role is played by the precise definition of the geodesic distance and that
one must carefully distinguish between boundary and initial conditions.
Such choices do indeed lead to different operators at the quantum level.
Nevertheless, it turned out that in our approximation scheme the anomalous dimension of the geodesic distance
is small and it is the same as that of the length of a arbitrary prescribed curve, considered in section \ref{sec:vol-hypersurfaces}.
We argued that this is due to the simple approximations that are employed in this exploratory work.
Finally, we also considered the Hausdorff dimension via the scaling of the volume of a geodesic ball
and argued that it coincides with the topological dimension.

Summarizing, we have considered various interesting geometric operators and computed their scaling properties
for the first time in the Asymptotic Safety scenario.
The results obtained in this work can be extended to more refined truncation schemes.
Two such directions are possible. 
The first consists in extending the ansatz for the EAA to more complex truncations.
The second is the introduction of mixing ans{\"a}tze which is essentially unexplored.
In the long run, such calculations will hopefully be useful to shed light on the possible connection
of the Asymptotic Safety scenario with other quantum gravity approaches, such as 
Causal Dynamical Triangulations \cite{Ambjorn:2012jv}.

\vspace{0.2cm}
\noindent
\subsubsection*{Acknowledgments}
The authors are grateful to Martin Reuter for many useful discussions. M.~B. is supported by DFG Grant RE 793/8-1.

\clearpage



\end{spacing}



\begin{thebibliography}{99}
%
\bibitem{W80}
S.~Weinberg, in \textit{General Relativity, an Einstein Centenary Survey}, S.~W.~Hawking and W.~Israel (Eds.), 
Cambridge University Press (1980) 790.
%
\bibitem{R98}
M.~Reuter, Phys.\ Rev.\ D \textbf{57} (1998) 971 and arXiv:hep-th/9605030.
%
\bibitem{W93}
C.~Wetterich, Phys.\ Lett.\ B \textbf{301} (1993) 90.
%
\bibitem{Pagani:2016dof} 
  C.~Pagani and M.~Reuter,
  Phys.\ Rev.\ D {\bf 95}, no. 6, 066002 (2017)
  [arXiv:1611.06522 [gr-qc]].
%
\bibitem{Pagani:2018mke} 
  C.~Pagani and M.~Reuter,
  JHEP {\bf 1807}, 039 (2018)
  [arXiv:1804.02162 [gr-qc]].
%
\bibitem{Lauscher:2005qz} 
  O.~Lauscher and M.~Reuter,
  JHEP {\bf 0510}, 050 (2005)
  [hep-th/0508202].
%
\bibitem{Reuter:2011ah} 
  M.~Reuter and F.~Saueressig,
  JHEP {\bf 1112}, 012 (2011)
  [arXiv:1110.5224 [hep-th]].
%
\bibitem{Carlip:2017eud} 
  S.~Carlip,
  Class.\ Quant.\ Grav.\  {\bf 34}, no. 19, 193001 (2017)
  doi:10.1088/1361-6382/aa8535
  [arXiv:1705.05417 [gr-qc]].
%
\bibitem{Ambjorn:1997di} 
  J.~Ambj{\o}rn, B.~Durhuus and T.~Jonsson,
  ``{\it Quantum Geometry : A Statistical Field Theory Approach}'',
  Cambridge University Press, Cambridge (UK), (1997)
%
\bibitem{Hamber:2009zz} 
  H.~W.~Hamber,
``{\it Quantum gravitation: The Feynman path integral approach}'',
Springer,  Berlin Germany (2009).
%
\bibitem{KPZ88}
A.~M.~Polyakov, Mod.\ Phys.\ Lett.\ A \textbf{02} (1987) 893;\\
V.~G.~Knizhnik, A.~M.~Polyakov and A.~B.~Zamolodchikov,\\ Mod.\ Phys.\ Lett.\ A \textbf{03} (1988) 819.  
%
\bibitem{DDK88}
F.~David, Mod.\ Phys.\ Lett.\ A \textbf{03} (1988) 1651;\\
J.~Distler and H.~Kawai, Nucl.\ Phys.\ B \textbf{321} (1989) 509.
%
\bibitem{Reuter:1996eg} 
  M.~Reuter and C.~Wetterich,
  Nucl.\ Phys.\ B {\bf 506}, 483 (1997)
  [hep-th/9605039].
%
\bibitem{Codello:2014wfa} 
  A.~Codello and G.~D'Odorico,
  Phys.\ Rev.\ D {\bf 92}, no. 2, 024026 (2015)
  [arXiv:1412.6837 [gr-qc]].
  %
\bibitem{Rechenberger:2012pm}
  S.~Rechenberger and F.~Saueressig,
  Phys.\ Rev.\ D {\bf 86} (2012) 024018,
  arXiv:1206.0657.
%
\bibitem{Calcagni:2013vsa} 
  G.~Calcagni, A.~Eichhorn and F.~Saueressig,
  Phys.\ Rev.\ D {\bf 87}, no. 12, 124028 (2013)
  [arXiv:1304.7247 [hep-th]].
%
\bibitem{Ellwanger:1993mw} 
  U.~Ellwanger,
  Z.\ Phys.\ C {\bf 62}, 503 (1994)
  [hep-ph/9308260].
%
\bibitem{Morris:1993qb} 
  T.~R.~Morris,
  Int.\ J.\ Mod.\ Phys.\ A {\bf 9}, 2411 (1994)
  [hep-ph/9308265].
%

\bibitem{Lauscher:2002sq}
  O.~Lauscher and M.~Reuter,
  Phys.\ Rev.\ D {\bf 66} (2002) 025026,
  hep-th/0205062.

\bibitem{Reuter:2002kd} 
M.~Reuter and F.~Saueressig,
Phys.\ Rev.\ D {\bf 66} (2002) 125001,
hep-th/0206145.

\bibitem{Codello:2007bd}
  A.~Codello, R.~Percacci and C.~Rahmede,
  Int.\ J.\ Mod.\ Phys.\ A {\bf 23} (2008) 143,
  arXiv:0705.1769.
  
\bibitem{Codello:2008vh} 
  A.~Codello, R.~Percacci and C.~Rahmede,
  Annals Phys.\  {\bf 324}, 414 (2009)
  [arXiv:0805.2909 [hep-th]].

\bibitem{Benedetti:2009rx}
  D.~Benedetti, P.~F.~Machado and F.~Saueressig,
  Mod.\ Phys.\ Lett.\ A {\bf 24} (2009) 2233,
  arXiv:0901.2984.
%
\bibitem{Benedetti:2010nr} 
D.~Benedetti, K.~Groh, P.~F.~Machado and F.~Saueressig,
JHEP {\bf 06} (2011) 079, 
arXiv:1012.3081.
%
\bibitem{Ohta:2013uca}
  N.~Ohta and R.~Percacci,
  Class.\ Quant.\ Grav.\  {\bf 31} (2014) 015024,
  arXiv:1308.3398.
%
\bibitem{Benedetti:2013jk}
  D.~Benedetti,
  EPL {\bf 102} (2013) 20007,
  arXiv:1301.4422.
%
\bibitem{Gies:2016con}
  H.~Gies, B.~Knorr, S.~Lippoldt and F.~Saueressig,
  Phys.\ Rev.\ Lett.\  {\bf 116} (2016) 211302,
  arXiv:1601.01800.
%
\bibitem{Falls:2014tra}
  K.~Falls, D.~F.~Litim, K.~Nikolakopoulos and C.~Rahmede,
  Phys.\ Rev.\ D {\bf 93} (2016) 104022,
  arXiv:1410.4815.
%
\bibitem{Ohta:2015efa}
  N.~Ohta, R.~Percacci and G.~P.~Vacca,
  Phys.\ Rev.\ D {\bf 92} (2015) 061501,
  arXiv:1507.00968.
%
\bibitem{Falls:2016wsa}
  K.~Falls, D.~F.~Litim, K.~Nikolakopoulos and C.~Rahmede,
  arXiv:1607.04962 [gr-qc].
%
\bibitem{Falls:2016msz}
  K.~Falls and N.~Ohta,
  Phys.\ Rev.\ D {\bf 94} (2016) 084005,
  arXiv:1607.08460.
%


\bibitem{Reuter:2008qx}
M.~Reuter and H.~Weyer,
Phys.\ Rev.\ D {\bf 80} (2009) 025001,
arXiv:0804.1475.

\bibitem{Benedetti:2012dx}
D.~Benedetti and F.~Caravelli,
JHEP {\bf 06} (2012) 017
Erratum: [JHEP {\bf 10} (2012) 157],
arXiv:1204.3541.

\bibitem{Demmel:2012ub}
  M.~Demmel, F.~Saueressig and O.~Zanusso,
  JHEP {\bf 11} (2012) 131,
  arXiv:1208.2038.

\bibitem{Dietz:2012ic}
J.~A.~Dietz and T.~R.~Morris,
JHEP {\bf 01} (2013) 108,
arXiv:1211.0955.

\bibitem{Bridle:2013sra}
I.~H.~Bridle, J.~A.~Dietz and T.~R.~Morris,
JHEP {\bf 03} (2014) 093,
arXiv:1312.2846.

\bibitem{Dietz:2013sba}
J.~A.~Dietz and T.~R.~Morris,
JHEP {\bf 07} (2013) 064,
arXiv:1306.1223.

\bibitem{Demmel:2014sga}
  M.~Demmel, F.~Saueressig and O.~Zanusso,
  JHEP {\bf 06} (2014) 026,
  arXiv:1401.5495.

\bibitem{Demmel:2014hla}
  M.~Demmel, F.~Saueressig and O.~Zanusso,
  Annals Phys.\  {\bf 359} (2015) 141,
  arXiv:1412.7207.

\bibitem{Demmel:2015oqa}
  M.~Demmel, F.~Saueressig and O.~Zanusso,
  JHEP {\bf 08} (2015) 113,
  arXiv:1504.07656.

\bibitem{Ohta:2015fcu}
  N.~Ohta, R.~Percacci and G.~P.~Vacca,
  Eur.\ Phys.\ J.\ C {\bf 76} (2016) 46,
  arXiv:1511.09393.

\bibitem{Labus:2016lkh}
P.~Labus, T.~R.~Morris and Z.~H.~Slade,
Phys.\ Rev.\ D {\bf 94} (2016) 024007,
arXiv:1603.04772.

\bibitem{Dietz:2016gzg}
J.~A.~Dietz, T.~R.~Morris and Z.~H.~Slade,
Phys.\ Rev.\ D {\bf 94} (2016) 124014,
arXiv:1605.07636.

\bibitem{Knorr:2017mhu}
B.~Knorr,
arXiv:1710.07055.

\bibitem{Falls:2017lst}
K.~G.~Falls, C.~S.~King, D.~F.~Litim, K.~Nikolakopoulos and C.~Rahmede,
arXiv:1801.00162.

\bibitem{Alkofer:2018fxj} 
  N.~Alkofer and F.~Saueressig,
  arXiv:1802.00498 [hep-th].
  
\bibitem{Manrique:2009uh}
  E.~Manrique and M.~Reuter,
  Annals Phys.\  {\bf 325} (2010) 785,
  arXiv:0907.2617.
%
\bibitem{Manrique:2010mq} 
  E.~Manrique, M.~Reuter and F.~Saueressig,
  Annals Phys.\  {\bf 326}, 440 (2011)
  [arXiv:1003.5129 [hep-th]].
%
\bibitem{Manrique:2010am}
  E.~Manrique, M.~Reuter and F.~Saueressig,
  Annals Phys.\  {\bf 326} (2011) 463,
  arXiv:1006.0099.
%
%
\bibitem{Becker:2014qya}
  D.~Becker and M.~Reuter,
  Annals Phys.\  {\bf 350} (2014) 225,
  arXiv:1404.4537.

\bibitem{Christiansen:2012rx}
  N.~Christiansen, D.~F.~Litim, J.~M.~Pawlowski and A.~Rodigast,
  Phys.\ Lett.\ B {\bf 728} (2014) 114,
  arXiv:1209.4038.
%
\bibitem{Codello:2013fpa}
  A.~Codello, G.~D'Odorico and C.~Pagani,
  Phys.\ Rev.\ D {\bf 89} (2014) 081701,
  arXiv:1304.4777.
%
\bibitem{Christiansen:2014raa}
  N.~Christiansen, B.~Knorr, J.~M.~Pawlowski and A.~Rodigast,
  Phys.\ Rev.\ D {\bf 93} (2016) 044036,
  arXiv:1403.1232.
%
\bibitem{Christiansen:2015rva}
  N.~Christiansen, B.~Knorr, J.~Meibohm, J.~M.~Pawlowski and M.~Reichert,
  Phys.\ Rev.\ D {\bf 92} (2015) 121501,
  arXiv:1506.07016.
%
\bibitem{Knorr:2017fus}
  B.~Knorr and S.~Lippoldt,
  Phys.\ Rev.\ D {\bf 96} (2017) 065020,
  arXiv:1707.01397.
  %
\bibitem{Christiansen:2017bsy} 
  N.~Christiansen, K.~Falls, J.~M.~Pawlowski and M.~Reichert,
  Phys.\ Rev.\ D {\bf 97}, no. 4, 046007 (2018)
  [arXiv:1711.09259 [hep-th]].
%
\bibitem{Eichhorn:2018akn} 
  A.~Eichhorn, P.~Labus, J.~M.~Pawlowski and M.~Reichert,
  arXiv:1804.00012 [hep-th].
  

\bibitem{Dou:1997fg}
  D.~Dou and R.~Percacci,
  Class.\ Quant.\ Grav.\  {\bf 15} (1998) 3449,
  hep-th/9707239.

\bibitem{Percacci:2002ie}
  R.~Percacci and D.~Perini,
  Phys.\ Rev.\ D {\bf 67} (2003) 081503,
  hep-th/0207033.

\bibitem{Dona:2013qba}
  P.~Don{\`a}, A.~Eichhorn and R.~Percacci,
  Phys.\ Rev.\ D {\bf 89} (2014) 084035,
  arXiv:1311.2898.

\bibitem{Christiansen:2017cxa} 
  N.~Christiansen, D.~F.~Litim, J.~M.~Pawlowski and M.~Reichert,
  Phys.\ Rev.\ D {\bf 97}, no. 10, 106012 (2018)
  [arXiv:1710.04669 [hep-th]].
  
\bibitem{Eichhorn:2018yfc} 
  A.~Eichhorn,
  arXiv:1810.07615 [hep-th].
  
  
\bibitem{deBrito:2018jxt} 
  G.~P.~De Brito, N.~Ohta, A.~D.~Pereira, A.~A.~Tomaz and M.~Yamada,
  Phys.\ Rev.\ D {\bf 98}, no. 2, 026027 (2018)
  [arXiv:1805.09656 [hep-th]].

\bibitem{Pawlowski:2005xe} 
  J.~M.~Pawlowski,
  Annals Phys.\  {\bf 322}, 2831 (2007)
  [hep-th/0512261].
%
\bibitem{Igarashi:2009tj} 
  Y.~Igarashi, K.~Itoh and H.~Sonoda,
  Prog.\ Theor.\ Phys.\ Suppl.\  {\bf 181}, 1 (2010)
  [arXiv:0909.0327 [hep-th]].
%
\bibitem{Pagani:2016pad} 
  C.~Pagani,
  Phys.\ Rev.\ D {\bf 94}, no. 4, 045001 (2016)
  [arXiv:1603.07250 [hep-th]].
\bibitem{Pagani:2017tdr} 
  C.~Pagani and H.~Sonoda,
  PTEP {\bf 2018}, no. 2, 023B02 (2018)
  [arXiv:1707.09138 [hep-th]].

\bibitem{Groh:2011dw} 
  K.~Groh, F.~Saueressig and O.~Zanusso,
  arXiv:1112.4856 [math-ph].

\bibitem{RSBook} 
M.~Reuter and F.~Saueressig,
  {\it Quantum Gravity and the Functional Renormalization Group -- The road towards Asymptotic Safety},
Cambridge University Press, in press.
\bibitem{Percacci:2017fkn} 
  R.~Percacci,
   {\it An Introduction to Covariant Quantum Gravity and Asymptotic Safety},
World Scientific, Singapore (2017).

\bibitem{Niedermaier:2006wt} 
  M.~Niedermaier and M.~Reuter,
  Living Rev.\ Rel.\  {\bf 9}, 5 (2006).

\bibitem{Litim:2001up} 
  D.~F.~Litim,
  Phys.\ Rev.\ D {\bf 64}, 105007 (2001)
  [hep-th/0103195].
  
\bibitem{Gray74}
A.~Gray, 
Michigan Math.~J.~20 (1974), no. 4, 329--344. 

\bibitem{Becker:2014pea} 
  D.~Becker and M.~Reuter,
  JHEP {\bf 1503}, 065 (2015)
  [arXiv:1412.0468 [hep-th]].


\bibitem{Daum:2010qt} 
  J.~E.~Daum and M.~Reuter,
  Phys.\ Lett.\ B {\bf 710}, 215 (2012)
  [arXiv:1012.4280 [hep-th]].
\bibitem{Daum:2013fu} 
  J.~E.~Daum and M.~Reuter,
  Annals Phys.\  {\bf 334}, 351 (2013)
  [arXiv:1301.5135 [hep-th]].
\bibitem{Pagani:2013fca} 
  C.~Pagani and R.~Percacci,
  Class.\ Quant.\ Grav.\  {\bf 31}, 115005 (2014)
  [arXiv:1312.7767 [hep-th]].
\bibitem{Pagani:2015ema} 
  C.~Pagani and R.~Percacci,
  Class.\ Quant.\ Grav.\  {\bf 32}, no. 19, 195019 (2015)
  [arXiv:1506.02882 [gr-qc]].
\bibitem{Reuter:2015rta} 
  M.~Reuter and G.~M.~Schollmeyer,
  Annals Phys.\  {\bf 367}, 125 (2016)
  [arXiv:1509.05041 [hep-th]].
\bibitem{Manrique:2011jc}
E.~Manrique, S.~Rechenberger and F.~Saueressig,
Phys.\ Rev.\ Lett.\  {\bf 106} (2011) 251302,
arXiv:1102.5012.

\bibitem{Rechenberger:2012dt}
S.~Rechenberger and F.~Saueressig,
JHEP {\bf 03} (2013) 010,
arXiv:1212.5114.
 
\bibitem{Biemans:2016rvp}
J.~Biemans, A.~Platania and F.~Saueressig,
Phys.\ Rev.\ D {\bf 95} (2017) 086013,
arXiv:1609.04813.

\bibitem{Biemans:2017zca}
J.~Biemans, A.~Platania and F.~Saueressig,
JHEP {\bf 05} (2017) 093,
arXiv:1702.06539.

\bibitem{Houthoff:2017oam}
W.~B.~Houthoff, A.~Kurov and F.~Saueressig,
Eur.\ Phys.\ J.\ C {\bf 77} (2017) 491,
arXiv:1705.01848.

\bibitem{Tsamis:1989yu} 
  N.~C.~Tsamis and R.~P.~Woodard,
  Annals Phys.\  {\bf 215}, 96 (1992).
\bibitem{Frob:2017apy} 
  M.~B.~Fr{\"o}b,
  Class.\ Quant.\ Grav.\  {\bf 35}, no. 3, 035005 (2018)
  [arXiv:1706.01891 [hep-th]].



\bibitem{Ambjorn:2012jv} 
  J.~Ambjorn, A.~Goerlich, J.~Jurkiewicz and R.~Loll,
  Phys.\ Rept.\  {\bf 519}, 127 (2012)
  doi:10.1016/j.physrep.2012.03.007
  [arXiv:1203.3591 [hep-th]].


  
\end{thebibliography}
\end{document}